\documentclass[a4paper,11pt]{article}
\usepackage{amsfonts}

\usepackage{graphicx}
\graphicspath{{Figures/}}

\usepackage{amsmath}
\usepackage{amssymb}
\usepackage{latexsym}
\usepackage[colorlinks]{hyperref}
\usepackage{color}
\usepackage{float}
\usepackage{cite}
\usepackage{makeidx}
\usepackage{colortbl}
\usepackage{csquotes}
\usepackage[squaren]{SIunits}
\usepackage{physics}

\usepackage[textfont={footnotesize,sf},labelfont={color=blue,bf,sf},labelsep=endash]{caption}
\usepackage[position=top,labelfont={color=blue,bf,sf}]{subfig}
\usepackage{bm}

\usepackage[title]{appendix}
\begin{document}
	\begin{titlepage}
		\setcounter{page}{1}
		\renewcommand{\thefootnote}{\fnsymbol{footnote}}
		\begin{flushright}
		\end{flushright}
		\begin{center}
			
			{\Large \bf 
	Tunneling Effect in Gapped	 Graphene Disk in Magnetic Flux\\ 
	and Electrostatic Potential
			}
			
			\vspace{5mm}
			
			{\bf A. Babe Cheikh}$^{a}$, {\bf A. Bouhlal}$^{b}$, 
			{\bf A. Jellal\footnote{\sf a.jellal@ucd.ac.ma}}$^{b,c}$
			and
			{\bf E. H. Atmani}$^{a}$
			
			\vspace{5mm}
			
			{$^{a}$\em Laboratory of Condensed Matter Physics
				and Renewed Energy, FST Mohammedia\\
				Hassan II University, Casablanca, Morocco}
			
			{$^{b}$\em Laboratory of Theoretical Physics,  
				Faculty of Sciences, Choua\"ib Doukkali University},\\
			{\em PO Box 20, 24000 El Jadida, Morocco}

			{$^{c}$\em Canadian Quantum  Research Center,
				204-3002 32 Ave Vernon, \\ BC V1T 2L7,  Canada}

			\vspace{3cm}
			\begin{abstract}
 We investigate the tunneling effect of a Corbino disk in  graphene in the presence of a variable magnetic flux $\Phi_{i}$ created by a solenoid piercing the inner disk under the effect of a finite mass term in the disk region $ (R_1< r<R_2) $ and an electrostatic potential. Considering  different regions, we explicitly determine the associated eigenspinors  in terms of Hankel functions. The use of matching conditions and asymptotic behavior of Hankel functions for large arguments, enables us to calculate  transmission and  other transport quantities. Our results show that the energy gap suppresses the tunneling effect by creating singularity points of zero transmission corresponding to the maximum shot noise peaks quantified by the Fano factor $ F $. The transmission as a function of the radii ratio $ R_2/R_1 $ becomes oscillatory with a decrease in periods and amplitudes. It can even reach one  (Klein tunneling) for large values of the energy gap.  The appearance of the minimal conductance at the points $ k_F R_1=R_1 \delta$ is observed.
 Finally we find that the electrostatic potential can control the effect of the band gap.
 
  \vspace{3cm}
 
 
	\end{abstract}

	\end{center}

	\end{titlepage}

	\section{Introduction}
Graphene consists of a single layer of carbon with one atom thick organized in a honeycomb structure, which was  isolated in 2004 by Novoselov and Geim \cite{Novoselov04}.
In the vicinity of the nodal points of high symmetry ($ K $ and $ K' $) of the first Bruillon zone, the electrons behave like Dirac fermions\cite{Divincenzo}
with a linear dispersion relation. Graphene is a semi-metal or a zero gap semiconductor in which charge carriers have a high mobility at room temperature\cite{Singh}. It has a unique chirality characteristic leading to several exotic transport factors such as the Klein tunnel effect\cite{Katsnelson}, anomalous quantum Hall effect\cite{McCann}, electron-hole symmetry\cite{Castroal} and many other effects. 
Graphene opened a piste toward for the discovery of different new materials in condensed matter physics and  
allows 
to have many applications in optoelectronics\cite{{Zhang05},{G2010}} as well as other areas.


On the other hand, a great attention was paid to
graphene quantum dots (QDs), which  are small fragments possessing electronic wavefunctions  confined in
disk \cite{b6}. 
Different techniques can be used to confine fermions in graphene
passing from 
magnetic fields \cite{199, 200} to cutting the flake into small nanostructures 
\cite{212, 222}.
Even with its interesting properties, unfortunately charge carrier confinement in graphene remains a challenge despite various methods. 
This is due to the  zero band gap in its energy spectrum and the manifestation  of the Klein tunneling effect. 
This means that electric current in graphene cannot be completely
shut off and such
 characteristic makes it unsuitable for the development of many
electronic devices.
%
This bear witness  to create a band gap in systems based on graphene.

A geometrically profile  was proposed by Rycerz
and  Suszalski
\cite{b8} to confine
fermions in graphene 
based on a Corbino disk  subjected to a solenoid magnetic potential. They investigated the transport properties by determining the transmission
and subsequently showed that 
 the conductance
as a
function of magnetic flux  exhibits periodic oscillations of the Aharonov-Bohm kind. Also it was 
found that such oscillations are
well-pronounced in the presence of  electrostatic potential, which breaks the cylindrical symmetry and
introduces the mode mixing.



 As matter of fact, 
 the creation of an energy gap remains a good choice, it is in this context that we subject the system considered in\cite{b8} to a mass term and study the tunneling effect.
 More precisely, 
 we analyze the influence of an energy gap created in the Corbino disk region in single-layer graphene ($ R_1 < r <R_2 $) pierced by a solenoid generating a magnetic flux $\Phi_{i}$ on the transmission probability, the Fano factor, the conductance and the magnitude of the conductance oscillations. 
As results, our  tunneling effect gets infected by the presence of 
the energy gap. Indeed, we show that  
the gap energy  leads to an increase of $ T_m $ for low doping accompanied by the appearance of singularities $ k_F R_1=R_1\delta $ of zero transmission.
Globally, it suppresses the tunneling effect by creating singularity points of zero transmission corresponding to the maximum shot noise peaks quantified by the Fano factor $ F $. We find that the presence of electrostatic potential breaks the symmetry and allows to control the effect of the energy gap for the cases where $ R_1 u_0\geqslant R_1\delta $.   

The paper is organized as follows. In section $ 2 $, we present  our theoretical study based on the solution of the Dirac equation in the different regions constituting our system. We use the continuity of the wave functions at the boundaries of the inner and outer disks together with the Hankel asymptotic solutions  for large arguments to calculate the transmission, conductance  and Fano factor. Section $ 3 $ is devoted to the discussion of our different numerical results. Finally, we conclude our work. 

	\section{Theoretical model}
\begin{figure}[H]\centering
	\includegraphics[width=0.3\linewidth]{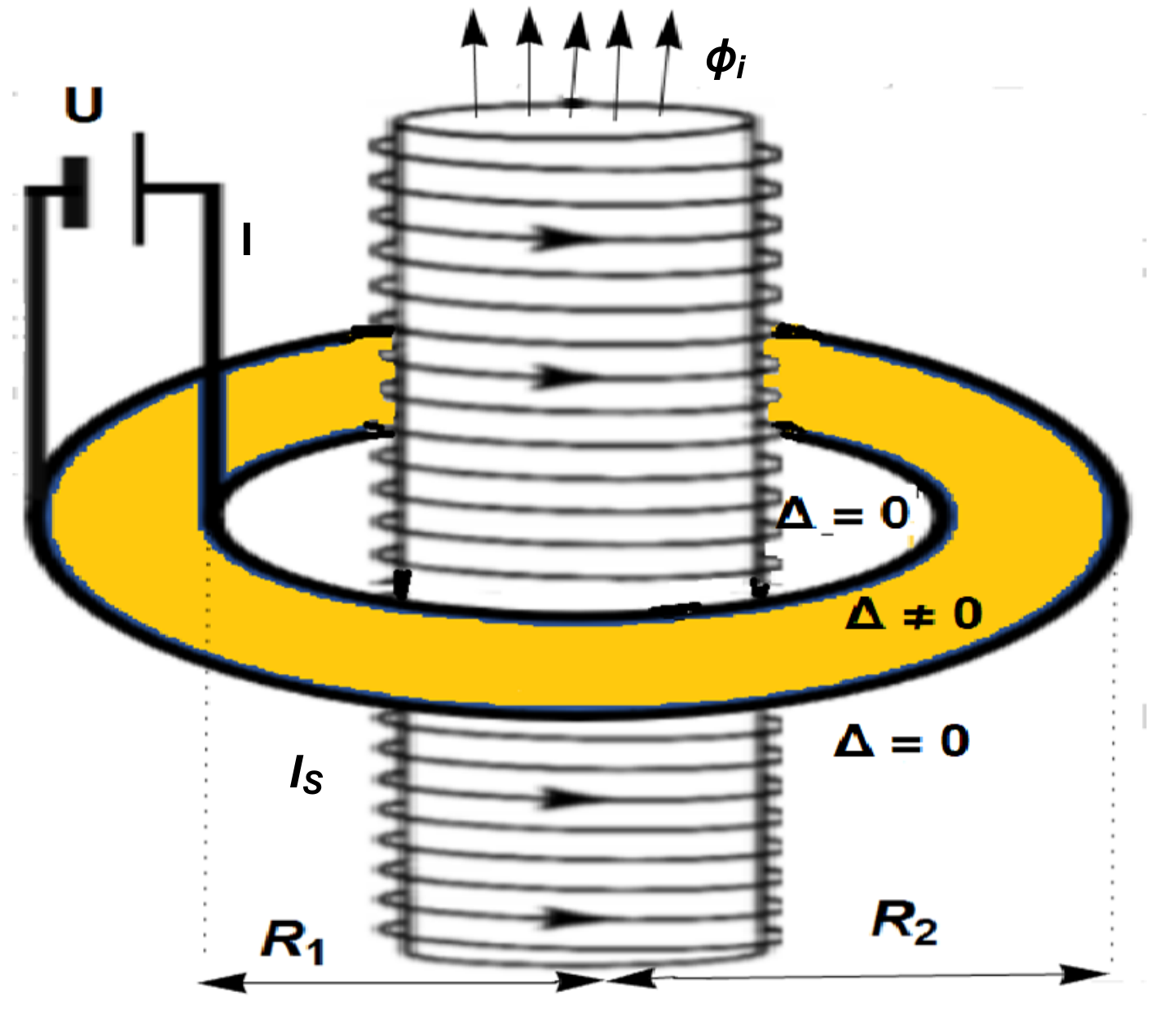}
	\caption{\sf (color online) the Corbino disk in graphene of the inner radius $R_1$ and the outer radius $R_2$, contacted by two electrodes. A separate gate electrode (not shown) allows the carrier concentration in the disk to be tuned around the neutrality point. A long solenoid, carrying the current $I_S$, generates the flux $\Phi_i$ piercing the inner disk area.}
	\label{fig 1}
\end{figure}

We consider an electron confined by an electrostatic potential in a Corbino disk in single-layer graphene and subjected to the effect of a mass term and a magnetic potential (see Fig. \ref{fig 1}), and then three diffusion regions are defined according to the values of the potential confinement given by
\begin{equation}\label{eq4}
	U(r)=
	\left\{%
	\begin{array}{ll}
		-U_0,  & \hbox{$R_1<r< R_2$} \\
		-U_{\infty}, &  \hbox{otherwise} \\
	\end{array}%
	\right.
\end{equation}

To achieve our task we introduce a mass term of the form
\begin{equation}\label{eq44}
	\Delta(r)=
	\left\{%
	\begin{array}{ll}
		\Delta,  & \hbox{$R_1<r< R_2$} \\
		0, &  \hbox{otherwise} \\
	\end{array}%
	\right.
\end{equation}
and consider the  vector potential in symmetric gauge is 
\begin{equation}\label{eq3}
	\vec A = \frac{\hbar}{e}\frac{\Phi_i}{\Phi_0 r} (-\sin\theta,\cos\theta)
\end{equation}  
Our system can be described by  the single-valley
Hamiltonian
\begin{equation}\label{eq2}
H=v_F (\vec p+e\vec{A})\cdot\vec \sigma+U(r) \mathbb{I}+\Delta\sigma_z
\end{equation}
where $v_F = 10^6$ m/s is the Fermi velocity, $ \vec{p}=(p_x,p_y) $ is the  momentum operator,
$ \vec{\sigma}=(\sigma_x,\sigma_y,\sigma_z) $ are Pauli  matrices in the basis of the two sublattices of $A$ and $B$ atoms.
Due to the symmetry of the system, we pass to the polar coordinates $ (r, \theta) $ and match the Hamiltonian (\ref{eq2}) as
\begin{equation}\label{eq5}
H= \begin{pmatrix}
U(r)+\Delta(r) & \partial_{-}\\ 
\partial_{+} &  U(r)-\Delta(r)
\end{pmatrix}
\end{equation}
where we use the notation 
\begin{equation}\label{eq6}
\partial_{\pm}=-i\hbar\nu_{F} e^{\pm i\theta}\left(\dfrac{\partial}{\partial r}\pm\frac{i}{r}\dfrac{\partial}{\partial\theta}\mp\frac{\Phi_i }{ \Phi_0 r}\right)
\end{equation}
and $ \Phi_0=h/e $ is the unit flux. 
Since the studied system has a cylindrical symmetry,  Hamiltonian (\ref{eq5}) commutes with the total angular momentum operator $J_z =L_z +S_z$. This implies that the eigenspinors
can be written as the product of a radial and angular function as
\begin{equation}\label{eq7}
\Psi_m(r,\theta)=\begin{pmatrix} \chi_{A}(r)\psi_m^+(\theta) \\ \chi_{B}(r)\psi_{m+1}^-(\theta)  \end{pmatrix}
\end{equation} 
such that 
\begin{equation}\label{eq8}
\psi_m^+(\theta)=\frac{e^{im\theta}}{\sqrt{2\pi}}\begin{pmatrix} 1  \\ 0  \end{pmatrix}, \qquad \psi_{m+1}^-(\theta)=\frac{e^{i(m+1)\theta}}{\sqrt{2\pi}}\begin{pmatrix} 0  \\ 1  \end{pmatrix}
\end{equation}
are  eigenstates of $J_z$ associated to the eigenvalues  $m\pm \frac{1}{2}$,
with the quantum numbers $m=0,\pm 1, \pm 2, \cdots$. 

To obtain the spinors we solve the famous Dirac equation in scattering problem $ H\Psi_m(r,\theta)=E \Psi_m(r,\theta) $ in the three regions shown in  Fig. \ref{fig 1}. In $ (r,\theta) $, the Dirac equation is now reduced to the radial form $ H_m(r)\chi_{m}(r) = E\chi_{m}(r) $ with $\chi_{m}=[\chi_{A},\chi_{B}]^T$ and 
\begin{equation}\label{eq9}
H_m(r)=
-i\hbar \nu_F \sigma_x \partial_{r} + U(r) + \Delta(r) + \hbar \nu_F \sigma_y
\begin{pmatrix}
\frac{m}{r}+\frac{\Phi_{i} }{\Phi_0 r} & 0\\ 
	0 &  \frac{m+1}{r}+\frac{\Phi_{i} }{\Phi_0 r}
\end{pmatrix}
\end{equation}
As the angular dependence of the wave function does not play a role for mode matching, then our analysis is effectively limited to the one-dimensional scattering problem for $ \chi_{m}(r) $ spinors. It is convenient, to assume
that the incident wave originates from the inner disk (outgoing wave propagates from $ r = \infty$, 
$ x\geq 0 $), the reflected wave entering the inner disk (incoming wave propagates from $ r = 0 $,  $ x\leq 0 $) and the transmitted wave is an outgoing wave.
By acting (\ref{eq9}) on $\chi_{m}$, we obtain
\begin{eqnarray}\label{eq10}
&&\left[\dfrac{\partial}{\partial r} +\frac{1}{r}\left(m+1+\frac{\Phi_{i} }{ \Phi_0}\right) \right]\chi_{B}(r)=i(\epsilon-u-\delta)\chi_{A}(r)
\\
&&
	\label{eq11}
\left[\dfrac{\partial}{\partial r} -\frac{1}{r}\left(m+\frac{\Phi_{i} }{\Phi_0}\right) \right]\chi_{A}(r)=i(\epsilon-u+\delta)\chi_{B}(r)
\end{eqnarray}
We can therefore write the following second order differential equation for $\chi_{A}(r)$ 
\begin{equation}\label{eq12}
\left[\rho^2\frac{\partial^2 }{\partial \rho^2}+\rho \dfrac{\partial}{\partial \rho} +\rho^2-\left(m+\frac{ \Phi_{i}}{\Phi_0}\right)^2\right]\chi_{A}(\rho)=0
\end{equation} 
which admits as solution the Hankel functions type $ H^{(1,2)}_\nu(\rho) $ where we put the variable $\rho=k r$ and two interesting quantities
\begin{equation}\label{eq13}
k =
\sqrt{\left|(\epsilon+u)^2-\delta^2\right|},\qquad
\nu =m+\frac{\Phi_{i}}{\Phi_0}
\end{equation}
with $\epsilon=\frac{E}{\hbar v_F}$, $u=\frac{U}{\hbar v_F}$ and  $\delta=\frac{\Delta}{\hbar v_F}$ are dimensionless parameters. Heavily doped graphene conductors are modeled by taking the limit of $  u (r)  
\rightarrow\pm\infty $ (hereinafter, the upper sign refers to the conduction
band and lower  to the valence band). In the case of electronic doping  ($  \epsilon-u>0 $) the outgoing and incoming normalized wave functions are given by 
\begin{equation}\label{eq14}
\chi^{out}_{m}=
\begin{pmatrix}
H^{(1)}_{\nu}(\rho)\\
i H^{(1)}_{\nu+1}(\rho)  
\end{pmatrix}, \quad
\chi^{inc}_{m}=
\begin{pmatrix}
	H_{\nu}^{(2)}(\rho)\\
	i H^{(2)}_{\nu+1}(\rho) 
\end{pmatrix}
\end{equation}
For the hole doping  $ (\epsilon-u <0) $ the wave functions are determined by using the conjugate expression of the spinors $ \tilde{\chi}^{out(inc)}_{m} =[\chi^{out(int)}]^*$. Let us take an interest in $  \epsilon-u > 0$, then the solution of (\ref{eq12}) can be written in each region.  Indeed, in the first region $r < R_1 $ and third one $ r > R_2 $, we have 
\begin{eqnarray}
	&&\label{eq15}
\chi^{(1)}_{m}=
\begin{pmatrix}
H^{(1)}_{\nu}(k_\infty r)\\
i H^{(1)}_{\nu+1}(k_\infty r)
\end{pmatrix}
+r_m
\begin{pmatrix}
H^{(2)}_{\nu}(k_\infty r)\\
i H^{(2)}_{\nu+1}(k_\infty r) 
\end{pmatrix}
\\
&&
\chi^{(3)}_{m}= t_m 
\begin{pmatrix}
H^{(1)}_{\nu}(k_\infty r)\\
i H^{(1)}_{\nu+1}(k_\infty r)
\end{pmatrix}
\end{eqnarray}
where the wave vector at infinity $ k_\infty=|\epsilon-u_\infty|\longrightarrow \pm\infty $.  
While
in the second region  (disk area $R_1<r<R_2$) it is
\begin{equation}\label{eq16}
	\chi^{(2)}_{m}= a
	\begin{pmatrix}
		H^{(1)}_{\nu}(k r)\\
		i H^{(1)}_{\nu+1}(k r) 
	\end{pmatrix}
	+b 
	\begin{pmatrix}
		H^{(2)}_{\nu}(k r)\\
		i H^{(2)}_{\nu+1}(k r)  
	\end{pmatrix}
\end{equation}
with the reflection $ r_m $ and transmission $ t_m $  coefficients, $ a $ and $ b $ are two constants of normalization.

 Using the asymptotic behavior of Hankel functions  $ H^{(\pm)}_\nu(\rho)\approx (2/\pi \rho)^{1/2} e^{\pm i(\rho-\nu\frac{\pi}{2}-\frac{\pi}{4})} $  for large arguments together with the relations $ H^{(1)}_{\nu+1}(\rho)= -i H^{(1)}_{\nu}(\rho)$ and 
$ H^{(2)}_{\nu+1}(\rho)= i H^{(2)}_{\nu}(\rho) $, we can simplify  (\ref{eq15}) to
\begin{eqnarray}
&&\chi^{(1)}_{m}=\frac{e^{+ik_{\infty} r}}{\sqrt{r}}
\begin{pmatrix}\label{eq17}
1\\
1
\end{pmatrix}
+r_{m} \frac{e^{-ik_{\infty} r}}{\sqrt{r}} 
\begin{pmatrix}
1\\
-1  
\end{pmatrix},\qquad r<R_1\\
&&\chi^{(3)}_{m}=t_{m}\frac{e^{+ik_{\infty} r}}{\sqrt{r}}
\begin{pmatrix}\label{eq18}
1\\
1
\end{pmatrix},\qquad r>R_2 
\end{eqnarray}
By solving the matching conditions 
\begin{equation}
	\chi^{(1)}_{m}(R_1)= \chi^{(2)}_{m}(R_1), \qquad \chi^{(2)}_{m}(R_2)= \chi^{(3)}_{m}(R_2) 
\end{equation}
  we find the transmission coefficient for the $m^{th}$ mode
\begin{equation}\label{eq19}
t_{m}=\frac{4e^{+i k_{\infty}(R_1-R_2)}}{\pi \sqrt{\left|(\epsilon+u)^2-\delta^2\right|R_1R_2}\left(\Gamma_{\nu}^{-}+i \Gamma_{\nu}^{+}\right)}
\end{equation}
giving rise to the transmission probability 
\begin{equation}\label{eq20}
T_m=\frac{16}{\pi^2 R_1R_2 \left|(\epsilon+u)^2-\delta^2\right|\left[(\Gamma_{\nu}^{+})^{2}+(\Gamma_{\nu}^{-})^{2}\right]}
\end{equation}
where we have set
\begin{equation}\label{eq21}
\Gamma_{\nu}^{+(-)}= \Im\left[H_{\nu}^{(1)}(k R_1) H_{\nu(\nu+1)}^{(2)}(k R_2)+(-)H_{\nu+1}^{(1)}(k R_1)H_{\nu+1(\nu)}^{(2)}(k R_2)\right]
\end{equation}

Two interesting physical quantities can be determined so far. Indeed,
the transmission  is used to calculate the linear response conductance by summing over the different modes according to the Landauer-Büttiker formula \cite{Buttiker1985}
 \begin{equation}\label{eq22}
 G=g_0\sum_m T_m
\end{equation} 
with  $ g_0=\frac{4 e^2}{h} $,  the  factor 4 accounts for the spin and valley degeneracy in graphene. 
The Fano factor quantifying the power of the shot noise for graphene, also results from the summation on the modes
 \begin{equation}\label{eq23}
 F=\frac{\sum_m T_m\left(1-T_m\right)}{\sum_m T_m}
 \end{equation}

\section{Results and discussions}
We study the effect of  energy gap $ R_1\delta $ created in the disk area (Fig. \ref{fig 1}) and the applied static potential $ U $ on the transmission probability $ T_m $, the Fano factor $ F $ and the conductance $ G $ as well as the magnitude of the conductance oscillations $ \Delta G $. Note that the effect of $U $
will be taken into account only in  Fig. \ref{fig 10} because our main task to study the impact of gap. In addition, our results will be presented by considering dimensionless  physical parameters.

\begin{figure}[H]\centering 
	\includegraphics[width=0.33\linewidth]{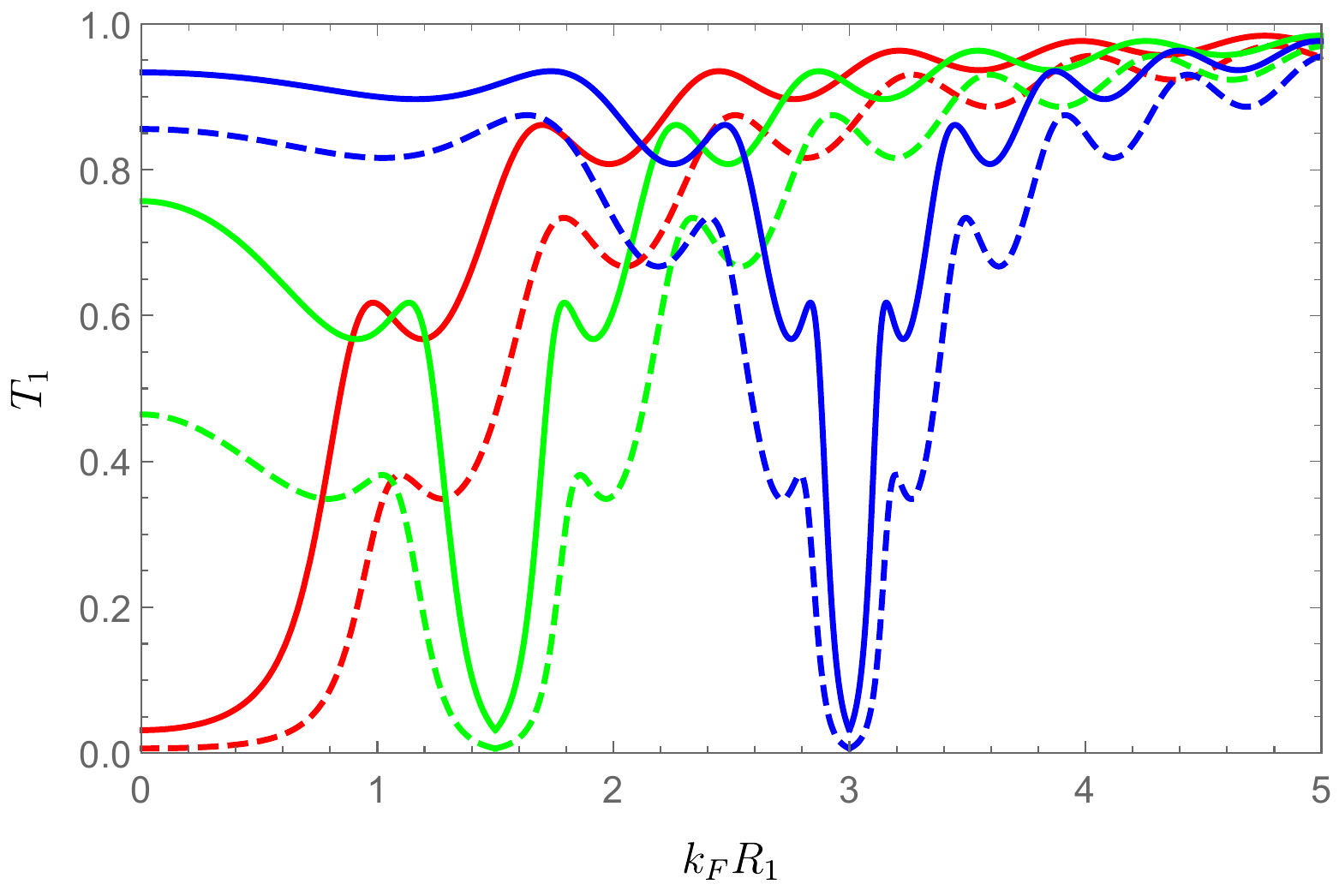},\includegraphics[width=0.33\linewidth]{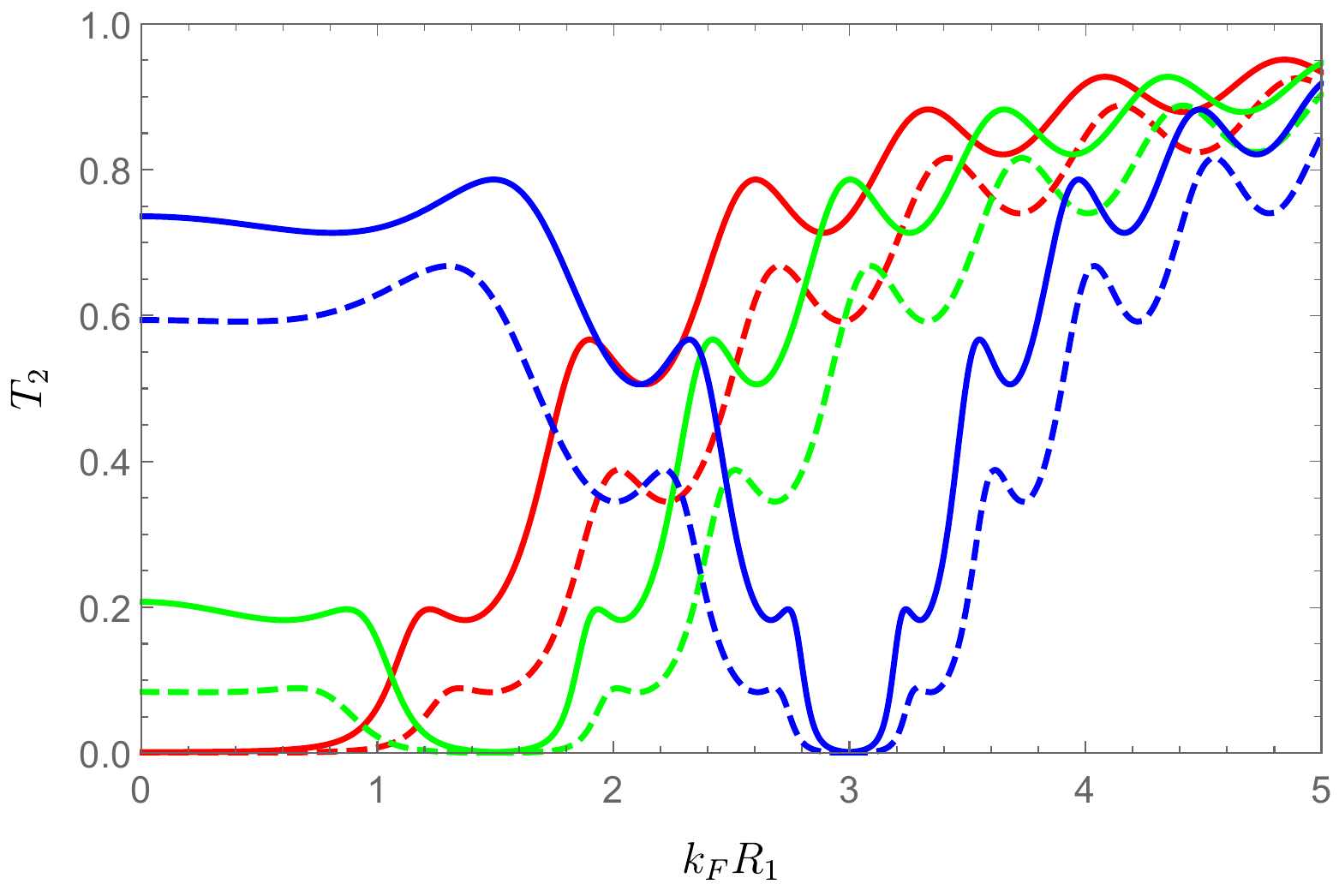},\includegraphics[width=0.33\linewidth]{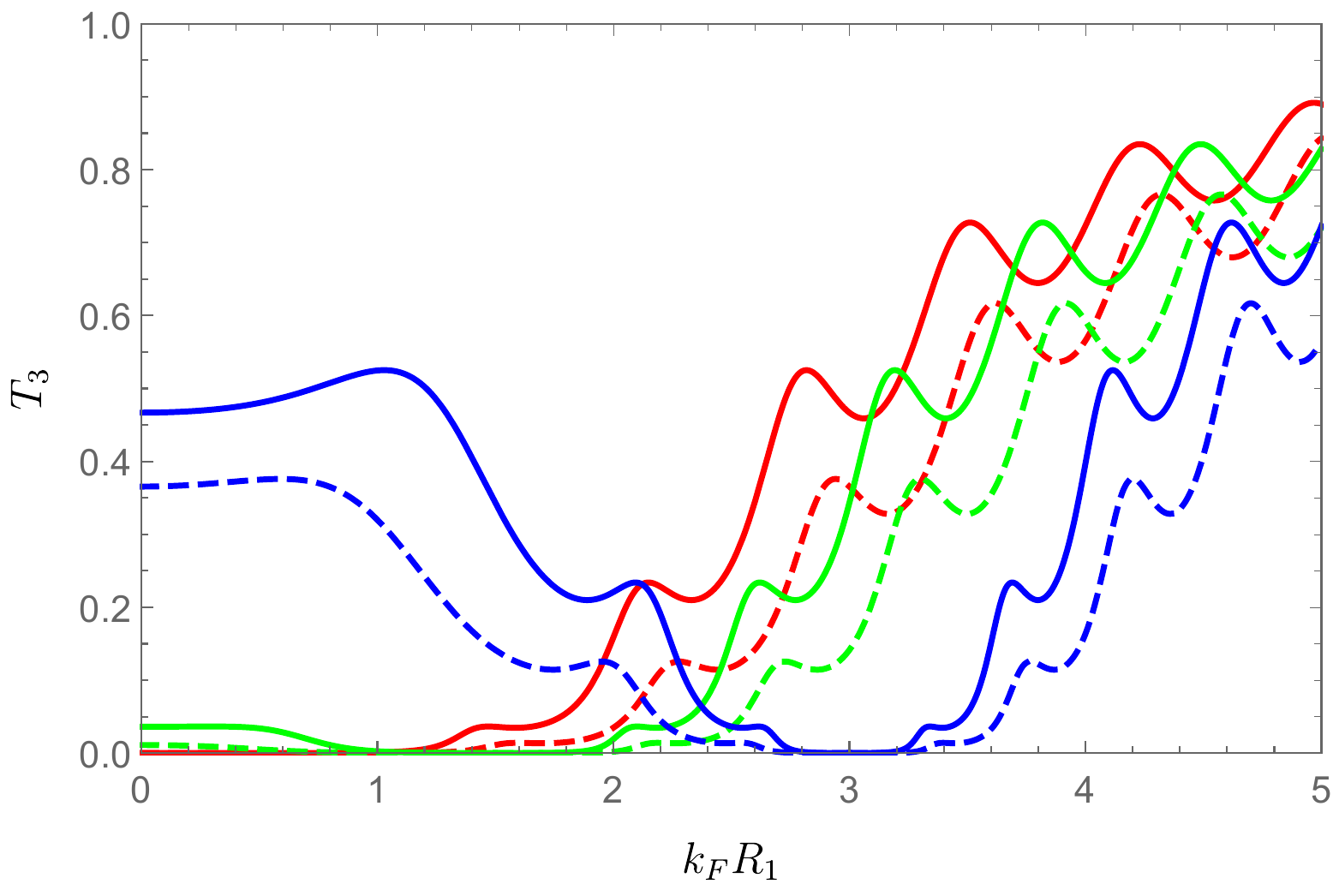}
	\caption{\sf (color online) The transmission $T_m$ ($m=1,2,3$) as a function of the doping $k_F R_1$ for the ratio radii fixed at $ R_2/R_1 = 5$ and different values of $ R_1 \delta$: 0 (red line), $ 1.5$ (green line), $ 3$ (blue line) with magnetic flux  $\Phi_i/\Phi_0=0$   (solid line) and $\Phi_i/\Phi_0=1/2$ (dashed line).}
	\label{fig 2}
\end{figure}

Fig. \ref{fig 2} shows the transmission probability $ T_m (m=1, 2, 3)$ as a function of doping $ k_F R_1 $ (with $ k_F $ is the Fermi wave number $ k_F = {|E|}/ {\hbar v_F} $) for three values of the energy gap $ R_1 \delta = 0 $ (red line), $ 1.5 $ (green line) and $ 3 $ (blue line) with magnetic flux (dashed line) and without magnetic flux (solid line). We notice that the inclusion of  energy gap $ R_1\delta $ in the graphene bands leads to an increase of $ T_m $ for low doping accompanied by the appearance of singularities $ k_F R_1=R_1\delta $ of zero transmission, then the transmission follows its progression as long as doping increases. This behavior 
decreases for the flux value
  $\Phi_i/\Phi_0=1/2$ by showing larger the bandwidths. According 3 panels, we observe that  our transmission decreases when the angular momentum $m$ increases.
\\

\begin{figure}[H]\centering
	\includegraphics[width=0.33\linewidth]{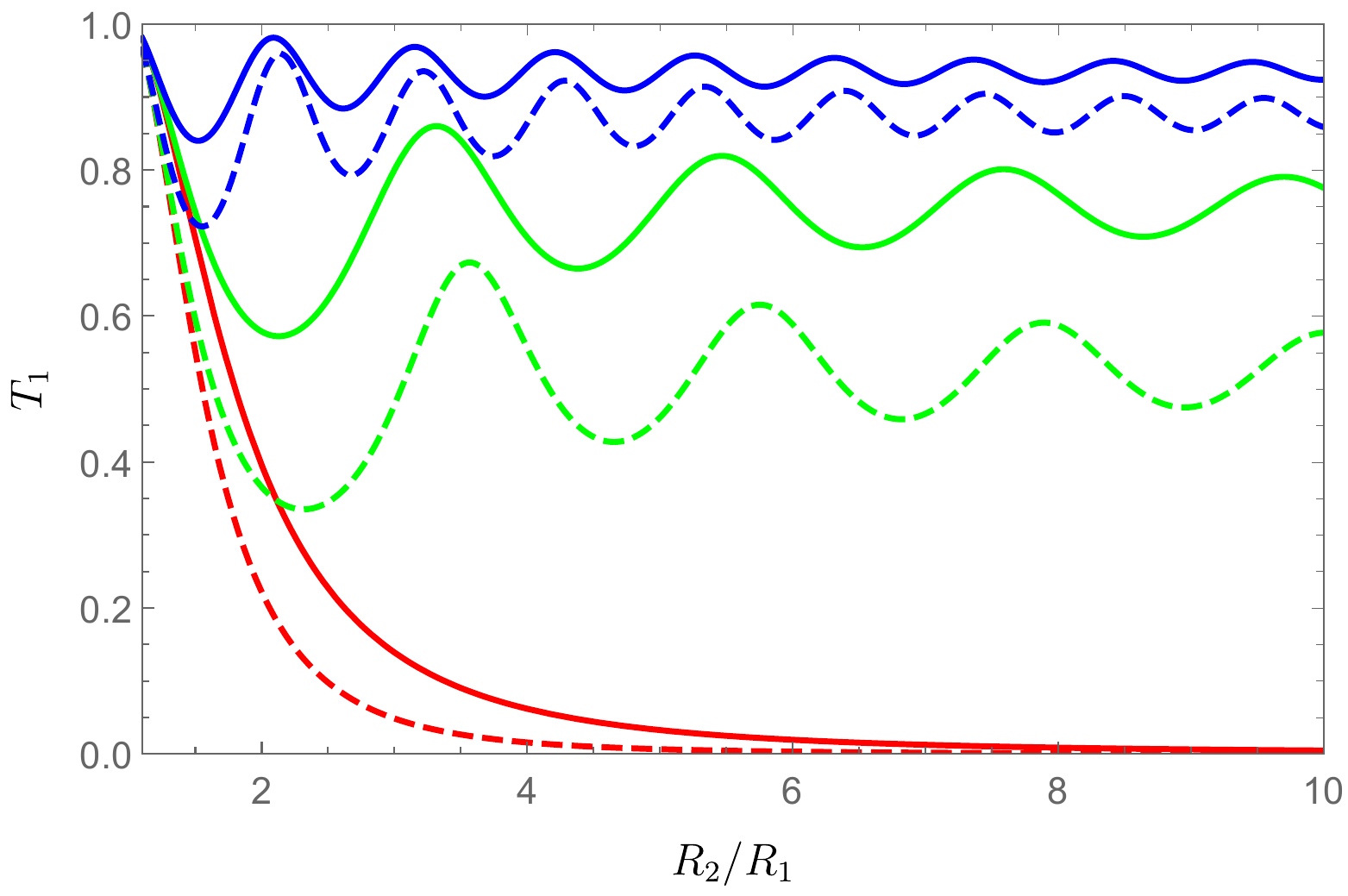},\includegraphics[width=0.33\linewidth]{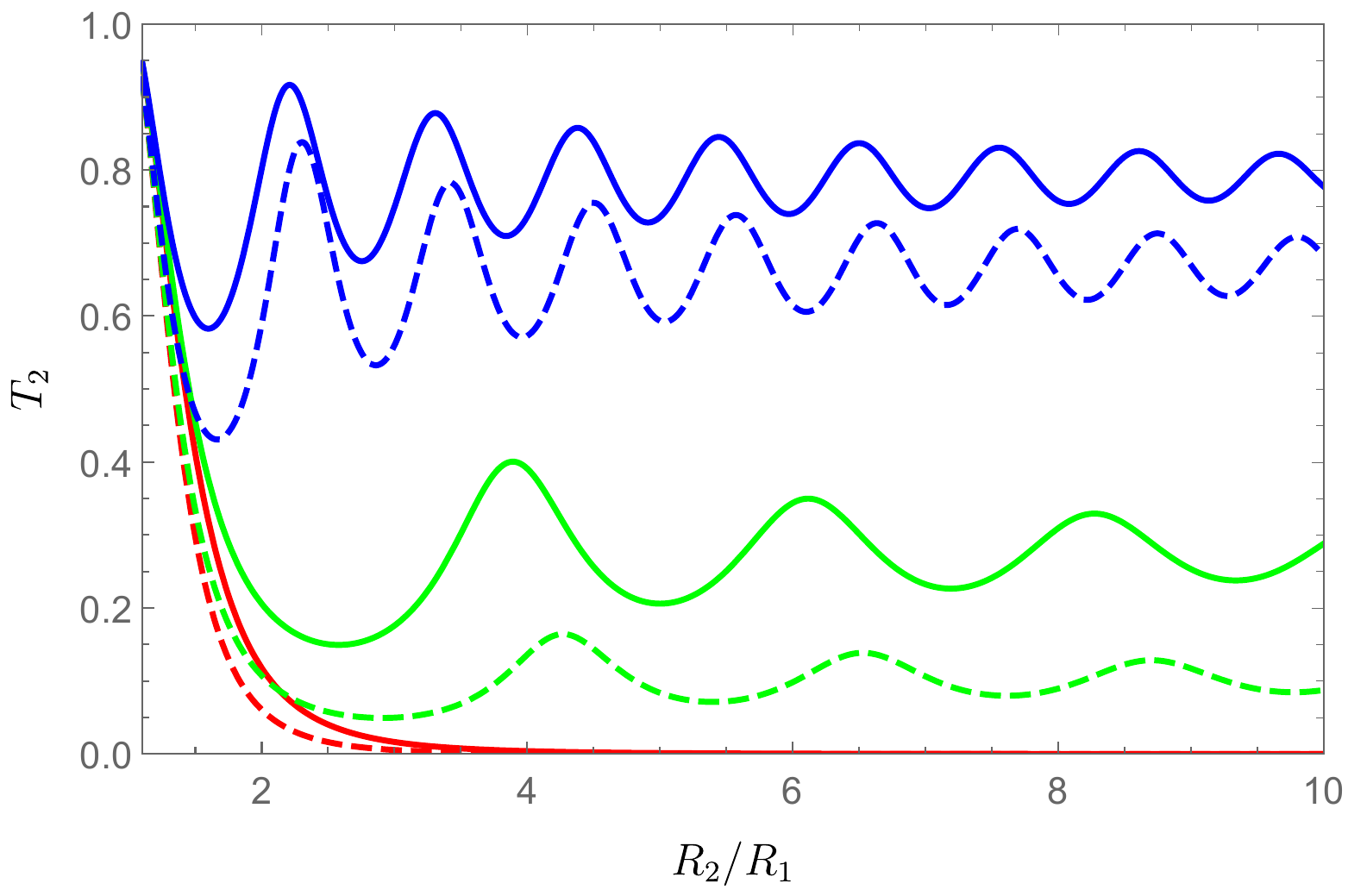},\includegraphics[width=0.33\linewidth]{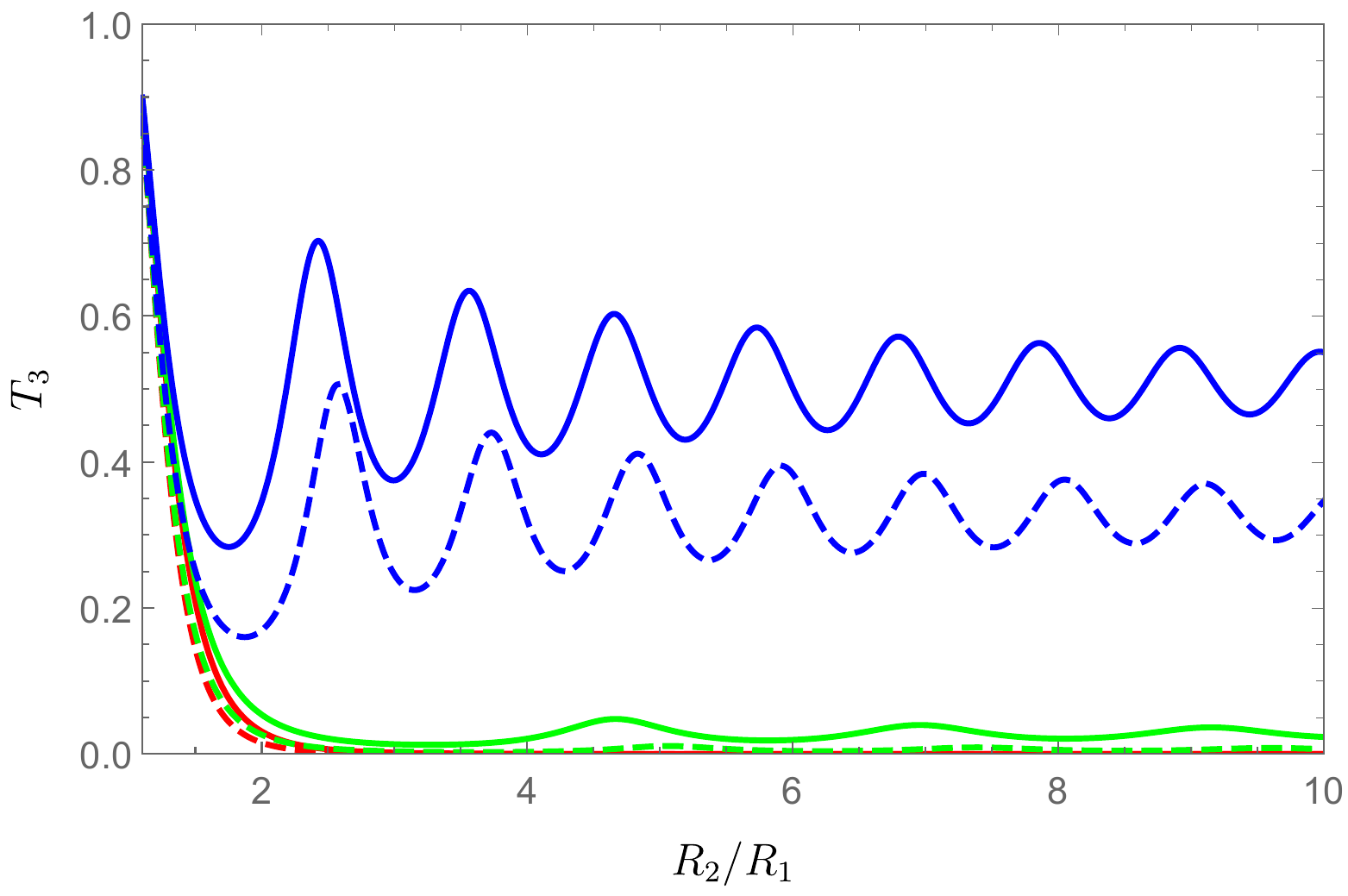}
	\caption{\sf (color online) The transmission  $T_m$ ($m=1,2,3$) as a function of the radii ratio $R_2/R_1$  for $k_F R_1=0.1$ and different values of $ R_1 \delta$: 0 (red line), $ 1.5 $ (green line), $ 3 $ (blue line) with magnetic flux $\Phi_i/\Phi_0=0$  (solid line) and $\Phi_i/\Phi_0=1/2$ (dashed line).}
	\label{fig 3}
\end{figure}
Fig. \ref{fig 3} presents the transmission probability $ T_m $ as a function of the radii ratio $ R_2 / R_1 $ and under the effect of three values of energy gap $ R_1 \delta = 0 $ (red line), 1.5 (line green), 3 (blue line), without 
(solid line) and with magnetic field (dashed line). The left panel corresponds to the value of the angular momentum $ m = 1 $, $ m = 2 $ (middle panel), $ m = $ 3 (right panel). For $ R_1 \delta = 0 $ (red line), Fig. \ref{fig 3} tells us that $ T_m $ decreases
exponentially toward zero as $ R_2 / R_1 $ increases. Now for non-zero gap (green, blue) we observe that $ T_m $ oscillates by increasing when $ R_1 \delta $ increases and even passes to a full transmission (Klein tunneling) for $ R_1 \delta = 3 $ (blue line) accompanied by a decrease in period and amplitude. The transmission decreases with increasing angular momentum.\\

 \begin{figure}[H]\centering
	\includegraphics[width=0.33\linewidth]{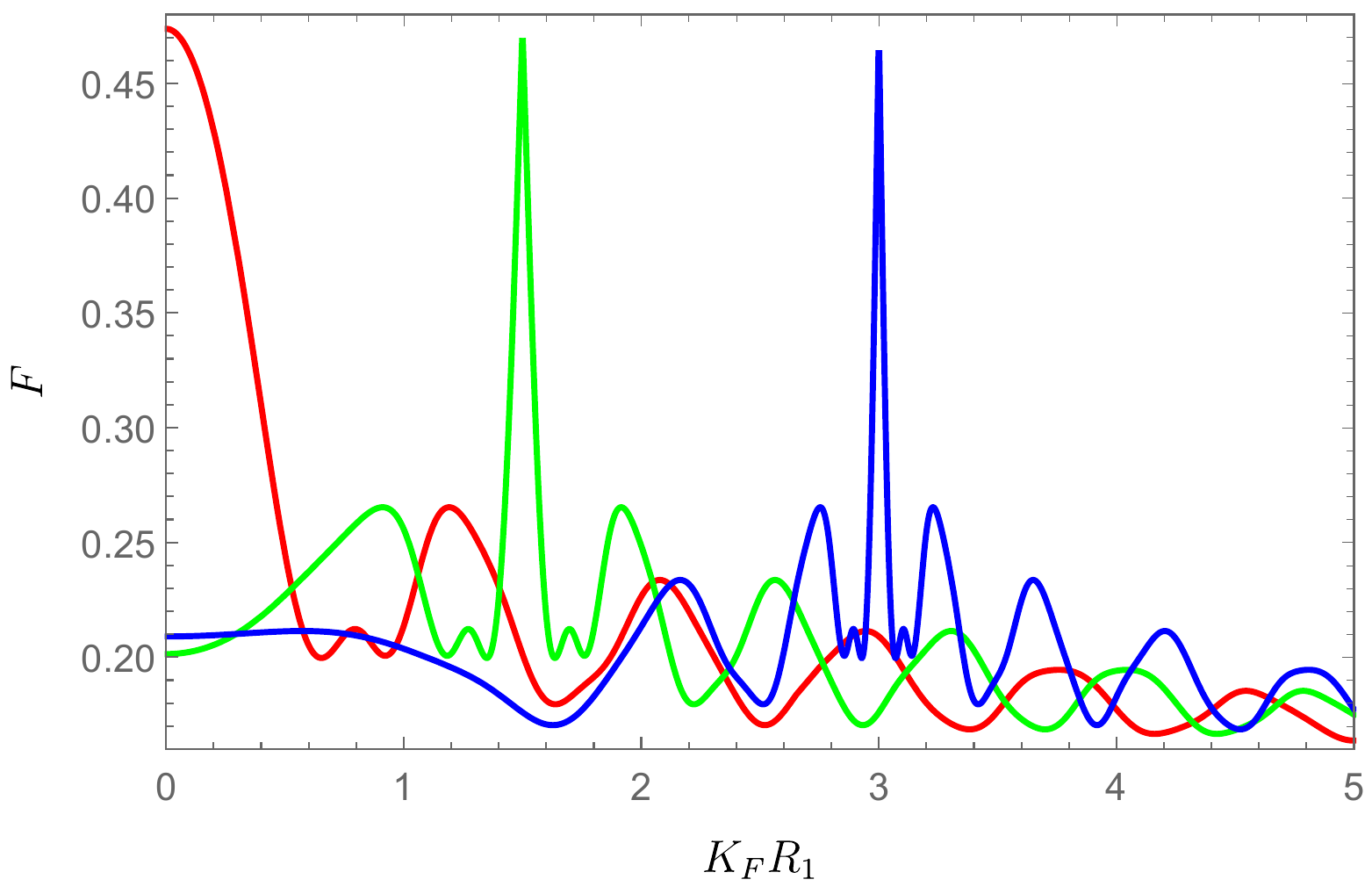}\ \ \ \ \ \ \ 
	\includegraphics[width=0.33\linewidth]{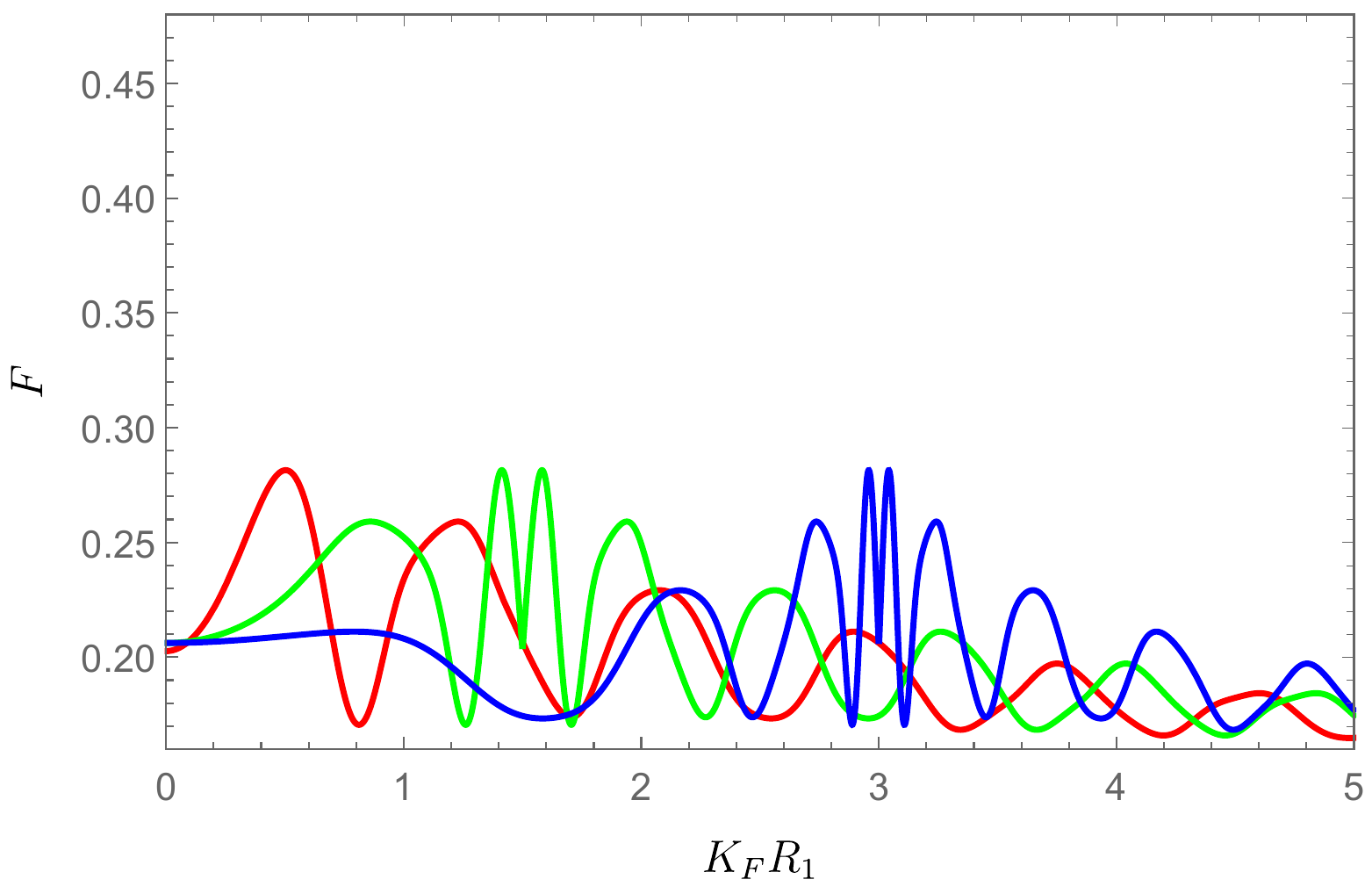}\\
	\caption{\sf (color online) The Fano factor $F$ as a function of the doping $k_F R_1$ for the radii ratio $R_2/R_1=5$ and different  values of energy gap $ R_1 \delta$: 0 (red line), 1.5 (green line), 3 (blue line). 
		Left panel:  $\Phi_i/\Phi_0=0$ and 
		right panel: $\Phi_i/\Phi_0=1/2$.}
	\label{fig 4}
\end{figure}
In Fig. \ref{fig 4} we plot the Fano factor $ F $ as a function of the doping $ k_F R_1 $ and under the effect of three values of energy gap $ R_1 \delta = 0 $ (red line), $ 1.5 $ (green line), $ 3 $ (blue line) with  magnetic flux (right panel)
and without  (left panel). For $ R_1 \delta = 0 $ (red line) and for a low doping $ k_F R_1 \rightarrow 0 $, we have a pseudo-diffusive regime $ F<1 $, which decreases and becomes oscillatory for high doping levels. For a non-zero energy gap (green, blue) and in the absence of magnetic flux we observe intense peaks at the points $ k_F R_1=R_1\delta $, then the curves follow an oscillatory process for high doping. In the presence of magnetic flux we observe a total disappearance of the peaks (right panel, green and blue line) and an appearance of peaks always identical and doubled at $ F = 0.28 $. The same peak appears for $ R_1 \delta = 0 $ (red line).\\

\begin{figure}[H]\centering
	\includegraphics[width=0.33\linewidth]{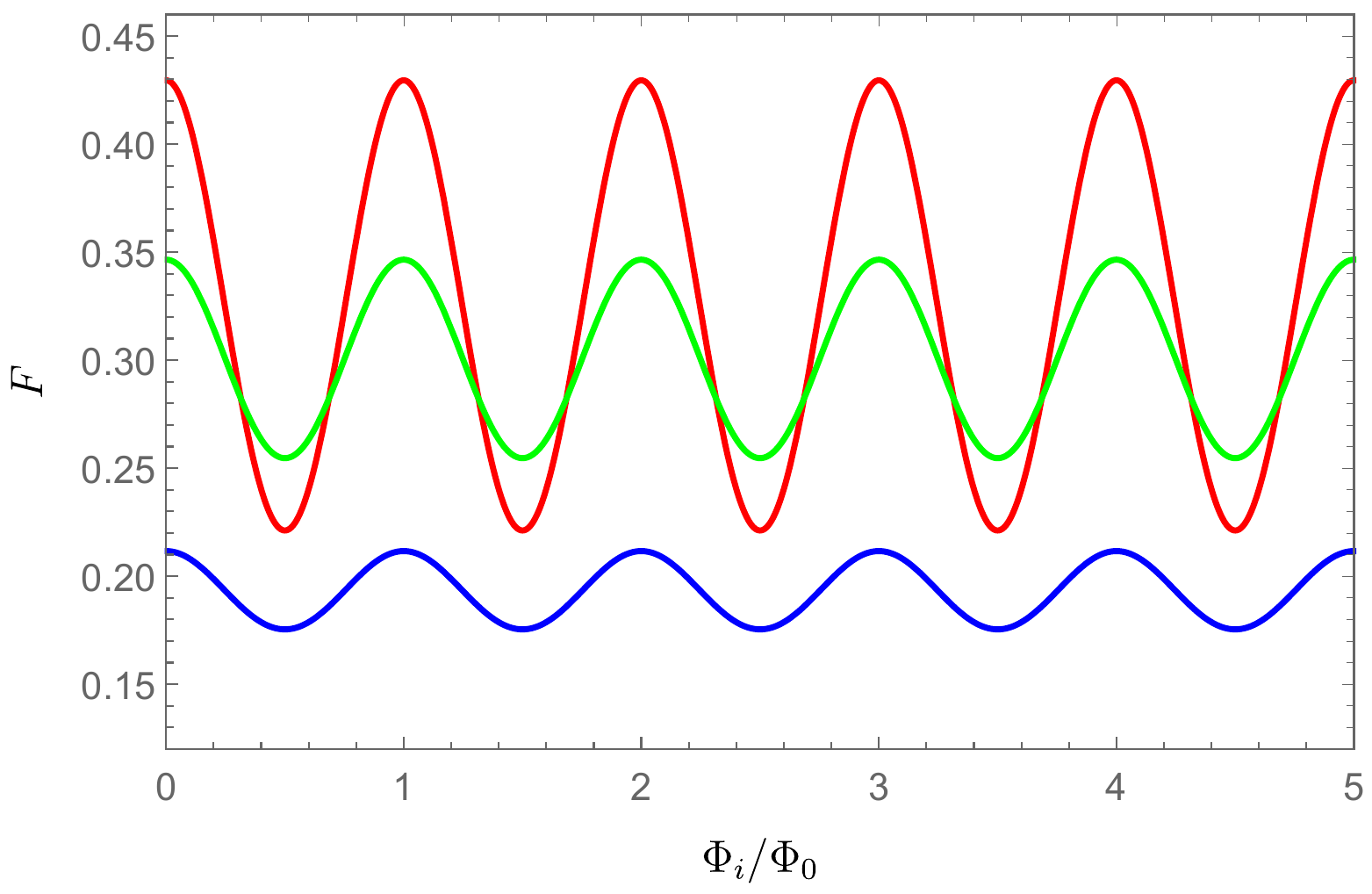}\ \ \ \ \ \ \
	\includegraphics[width=0.33\linewidth]{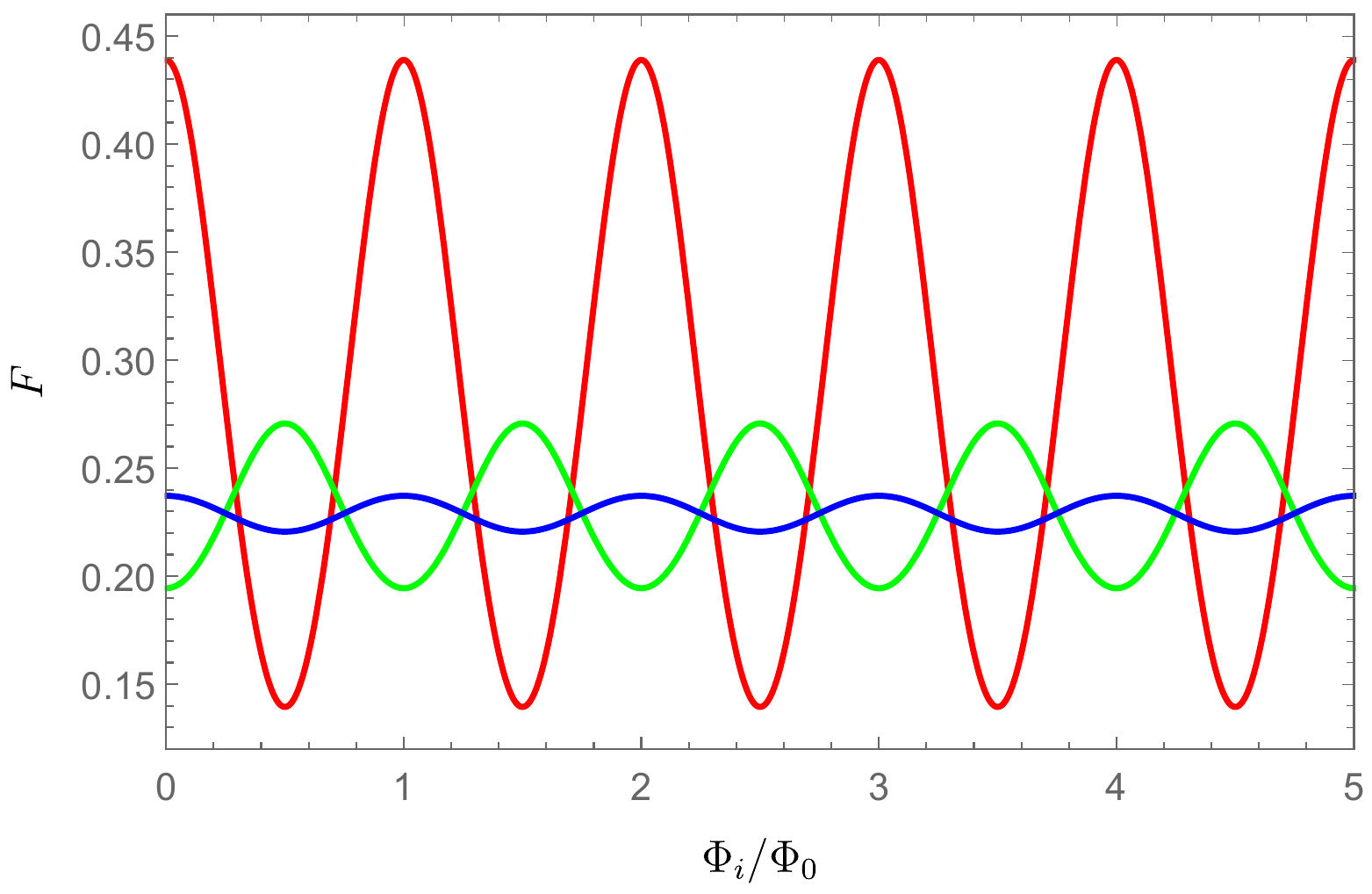}
	\caption{\sf (color online) The Fano factor $F$ as a function of the flux piercing the inner disk area $\Phi_i/\Phi_0$ for a doping fixed at $ k_F R_1 = 0.2 $ and different values of  $ R_1 \delta$: 0 (red line), $ 0.4 $ (green line), $ 0.8 $ (blue line) with the ratio  $ R_2 / R_1 = 5 $ (left panel) and $ 10 $ (right panel).}
	\label{fig 5}
\end{figure}
In Fig. \ref{fig 5} we plot the Fano factor $ F $ as a function of the magnetic flux $ \Phi_i/\Phi_0 $ piercing the inner disk, in the presence of three values of energy gap $ R_1 \delta = 0 $ (red line), $ 0.4 $ (green line) , $ 0.8 $ (blue line). In the left panel therefore we see that the shot noise presents a periodic oscillation depending on the flux around the value of amplitude $ F = 0.43 $. We notice that  the amplitude of these oscillations decreases by increasing the energy gap and the noise becomes $ F = 0.346 $ (green line) for $ R_1 \delta = 0.4 $, $ F = 0.21 $ (blue line) for $ R_1 \delta = 0.4 $. In the right panel corresponds to $ R_2 / R_1=10$ we observe a phase shift a new increase in noise by increasing the energy gap.\\

\begin{figure}[H]\centering
	\includegraphics[width=0.33\linewidth]{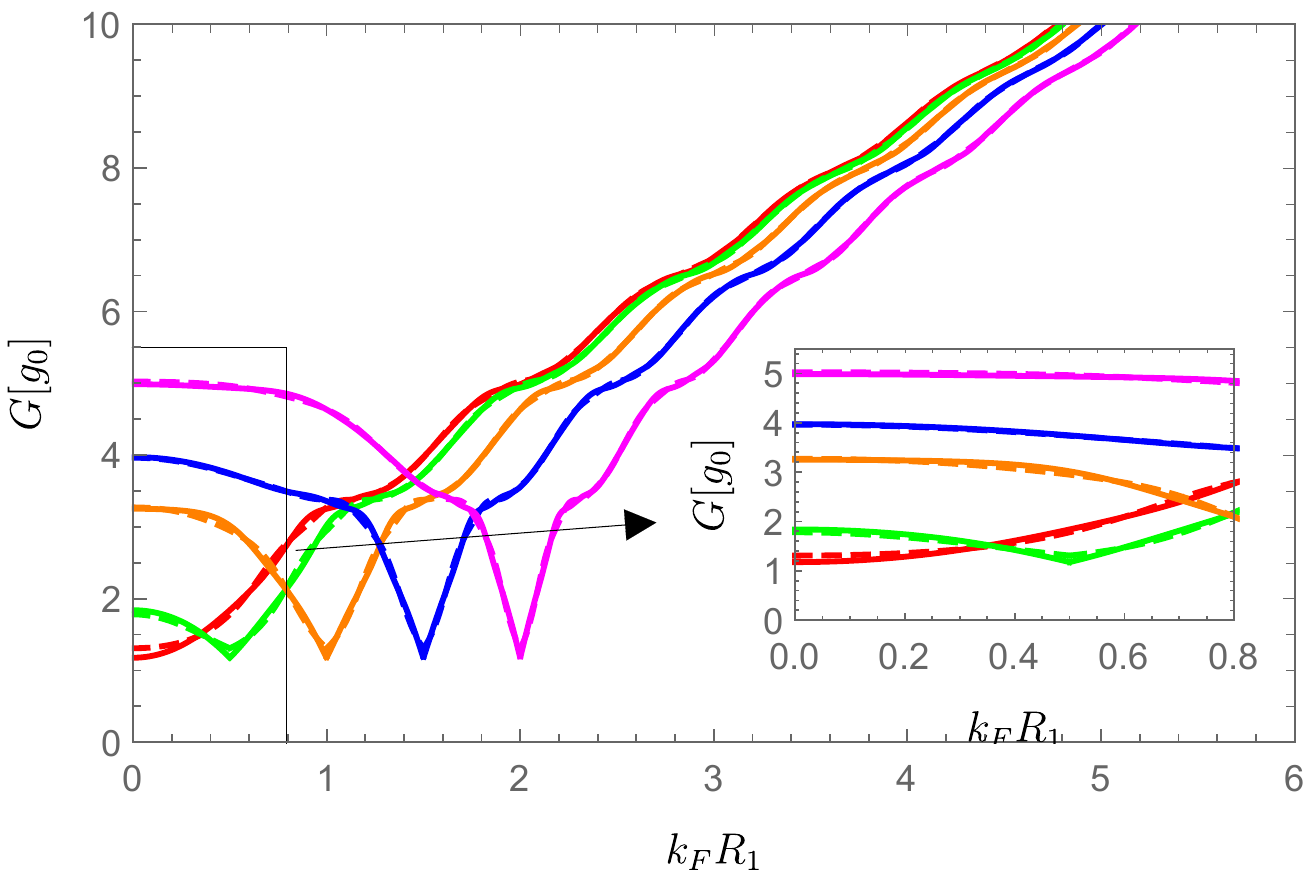}\ \ \ \ \ \ \
	\includegraphics[width=0.33\linewidth]{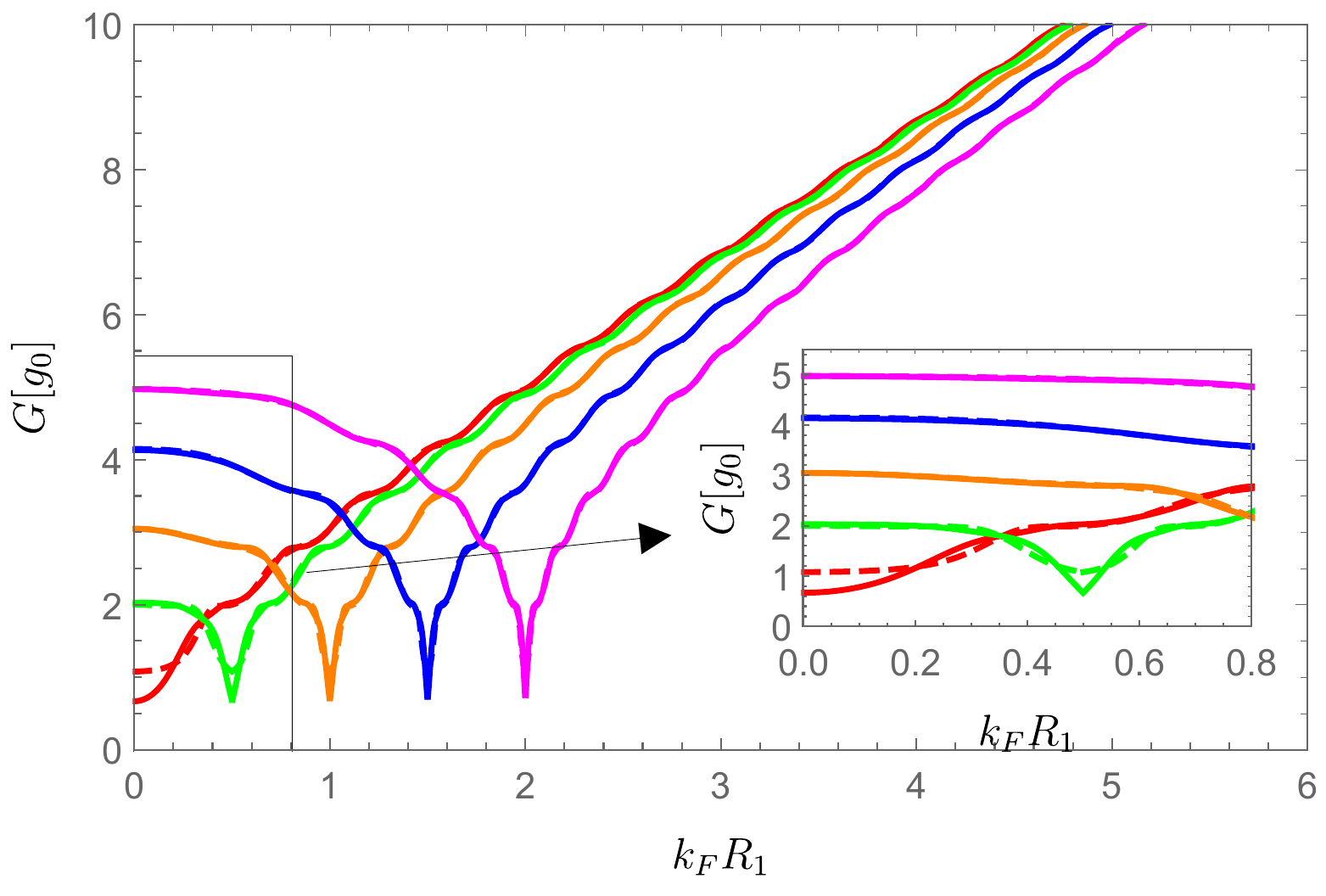}\\
	\caption{\sf (color online) The conductance $G$ 
		as a function of the doping $k_F R_1$ for different values of energy gap  $ R_1 \delta$: (red line), 0.5 (green line), 1 (orange line), 1.5 (blue line), 2 (magenta line) and for two values of the flux piercing the inner disk area $\Phi_i/\Phi_0=0$ (solid line), $\Phi_i/\Phi_0=1/2$ (dashed line) with the radii ratio  $ R_2/R_1 =5 $ (left panel) and  $ R_2/R_1 =10 $ (right panel). }
	\label{fig 6}
\end{figure}

Fig. \ref{fig 6} shows that the conductance can be modulated by the doping $ k_F R_1 $. By using the Hankel functions  properties in  (\ref{eq22}), it can be approximated linearly by $ G \approx 2g_0 k_F R_1 $. Then, it is clearly seen that as the doping increases, the conductance increases as well. This result is valid in the absence of energy gap, i.e.  $ R_1 \delta = 0 $. Now for zero doping, the conductance increases by increasing the energy gap and becomes minimal representing singularities in $ k_F R_1=R_1 \delta $ then it follows the same aspect as gapless case $ G (R_1 \delta = 0) $ studied in \cite{Rycerz2020}. It is important to note that the effect of magnetic flux also decreases by increasing energy gap. \\

\begin{figure}[H]\centering
	\includegraphics[width=0.33\linewidth]{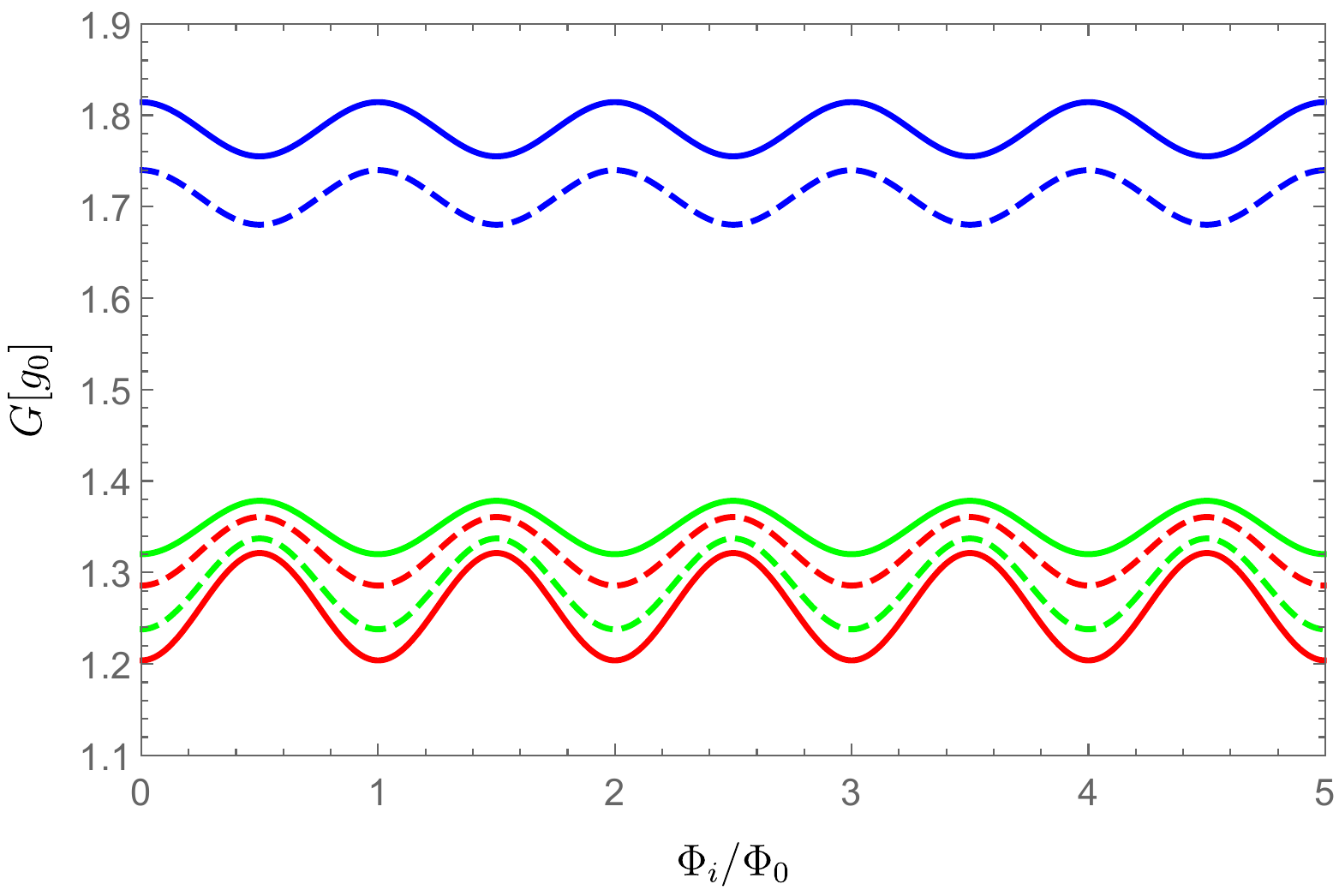}\ \ \ \ \ \ \
	\includegraphics[width=0.33\linewidth]{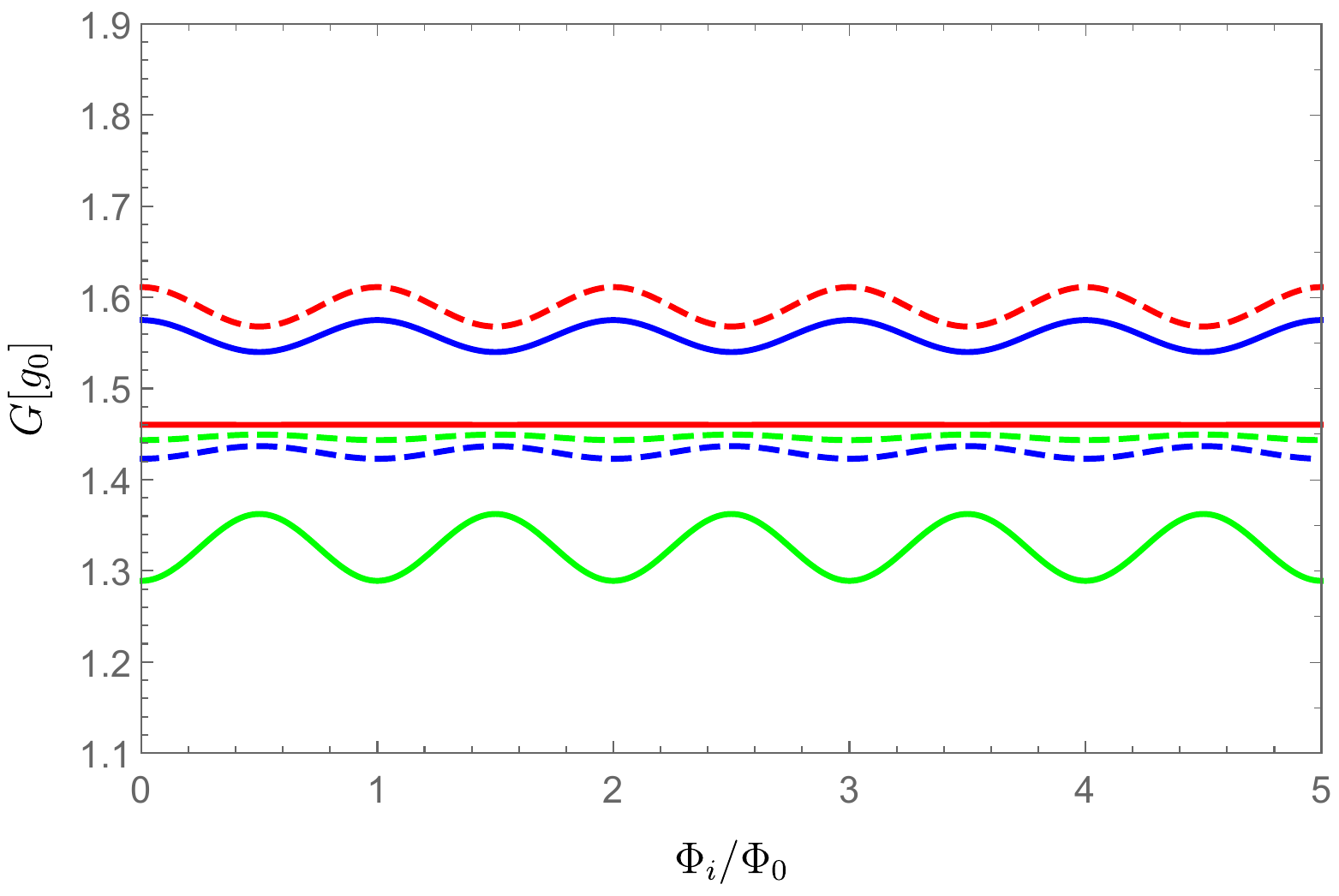}
	\caption{\sf (color online)   The conductance $ G $ as a function of the flux piercing the inner disk area for radii ratio $R_2/R_1 = 5$ and different values of gap energy $ R_1 \delta$: 0 (red line),
		$ 0.25 $ (green line), $ 0.5 $ (blue line). In the left panel we have plotted for two values of the doping $k_F R_1=0.1$ (solid line), $0.2$ (dashed line), and in the right panel we have plotted for $ k_F R_1=0.322 $ (solid line), $ 0.4 $ (dashed line). }
	\label{fig 7}
\end{figure} 

 In  Fig. \ref{fig 7} we present the conductance as a function of the flux piercing the inner disk under suitable conditions of the physical parameters. Indeed, let us notice first for zero doping limit the transmission (\ref{eq20}) can be simplified 
 to
 \cite{Buttiker1985} 
 \begin{equation}\label{eq1}
 T_m = \frac{1}{\cosh^2[\ln(R_2/R_1)(m+\Phi_d/\Phi_0)]}
 \end{equation} 
 and therefore  the conductance  (\ref{eq22}) becomes 
 \begin{equation}\label{eq24}
 	G=g_0\sum_m T_m(k_F\rightarrow 0)=\sum^{\infty}_{n=0} G_n \cos(\frac{2 \pi \Phi_i}{\Phi_0}) 
 \end{equation}
 where the involved quantities are given by 
 \begin{equation}
 	G_0=\frac{2g_0}{\ln(R_2/R_1)}, \qquad G_n=\frac{4\pi^2 (-)^n
 		n g_0}{\ln(R_2/R_1)^2\sinh[\pi^2 n/\ln(R_2/R_1)]}
 \end{equation}
It is clear that the expressions (\ref{eq13}), (\ref {eq20}), (\ref {eq22}) and (\ref {eq24}) show a perfectly periodic functional dependence of $ G $ on $ \Phi_i /\Phi_0 $ 
with an average value $ G_0 $ equal to the pseudo-diffusion conductance. Now by introducing an energy gap we observe in Fig. \ref{fig 7} a coincidence of periods followed by a decrease in the amplitudes of conductance. Additionally, we notice that $G$ increases for $ k_F R_1 < R_1\delta $ and but decreases for $ k_F R_1 > R_1\delta $.
 
\begin{figure}[H]\centering
	\includegraphics[width=0.32\linewidth]{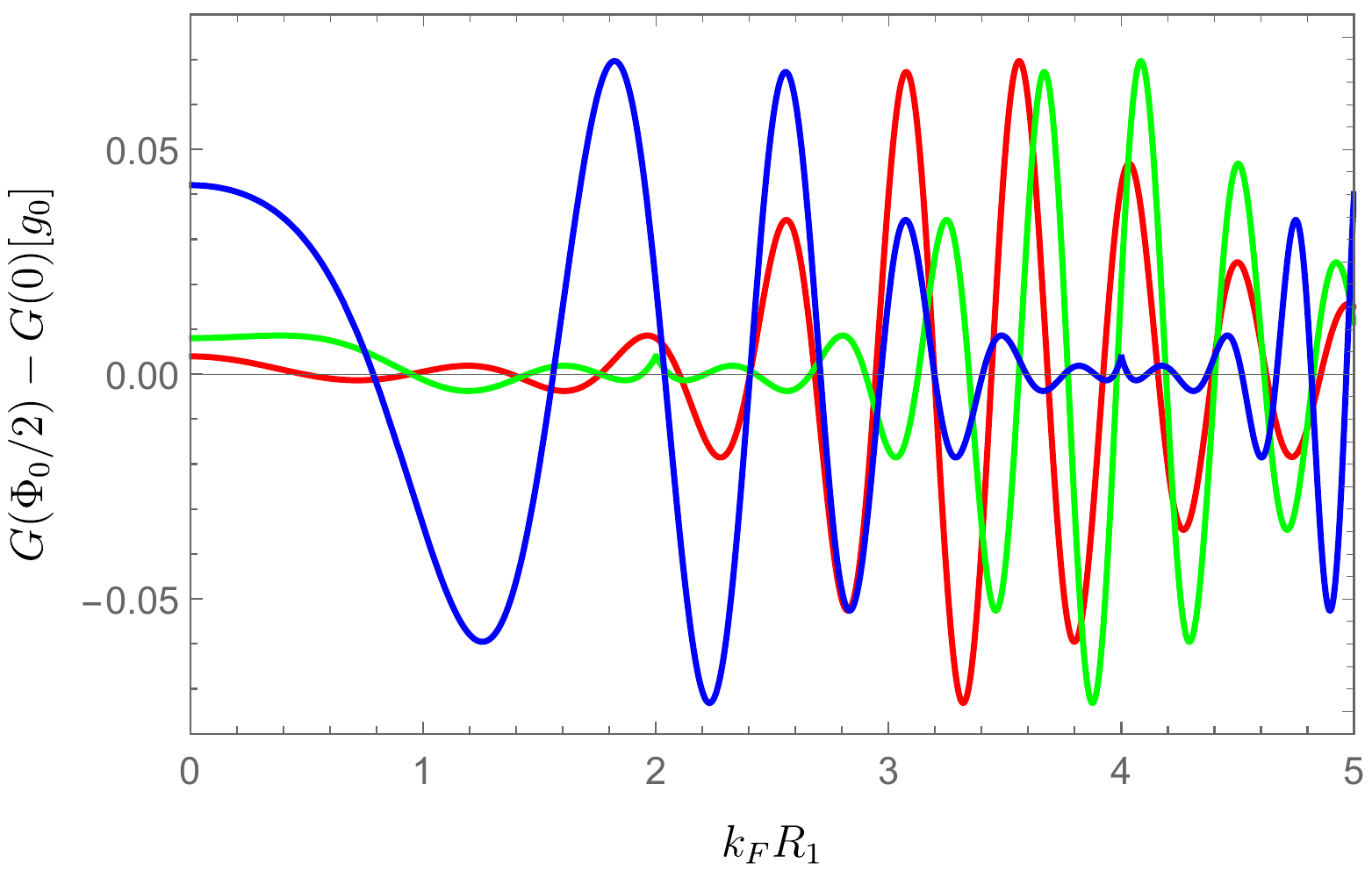}
	\includegraphics[width=0.32\linewidth]{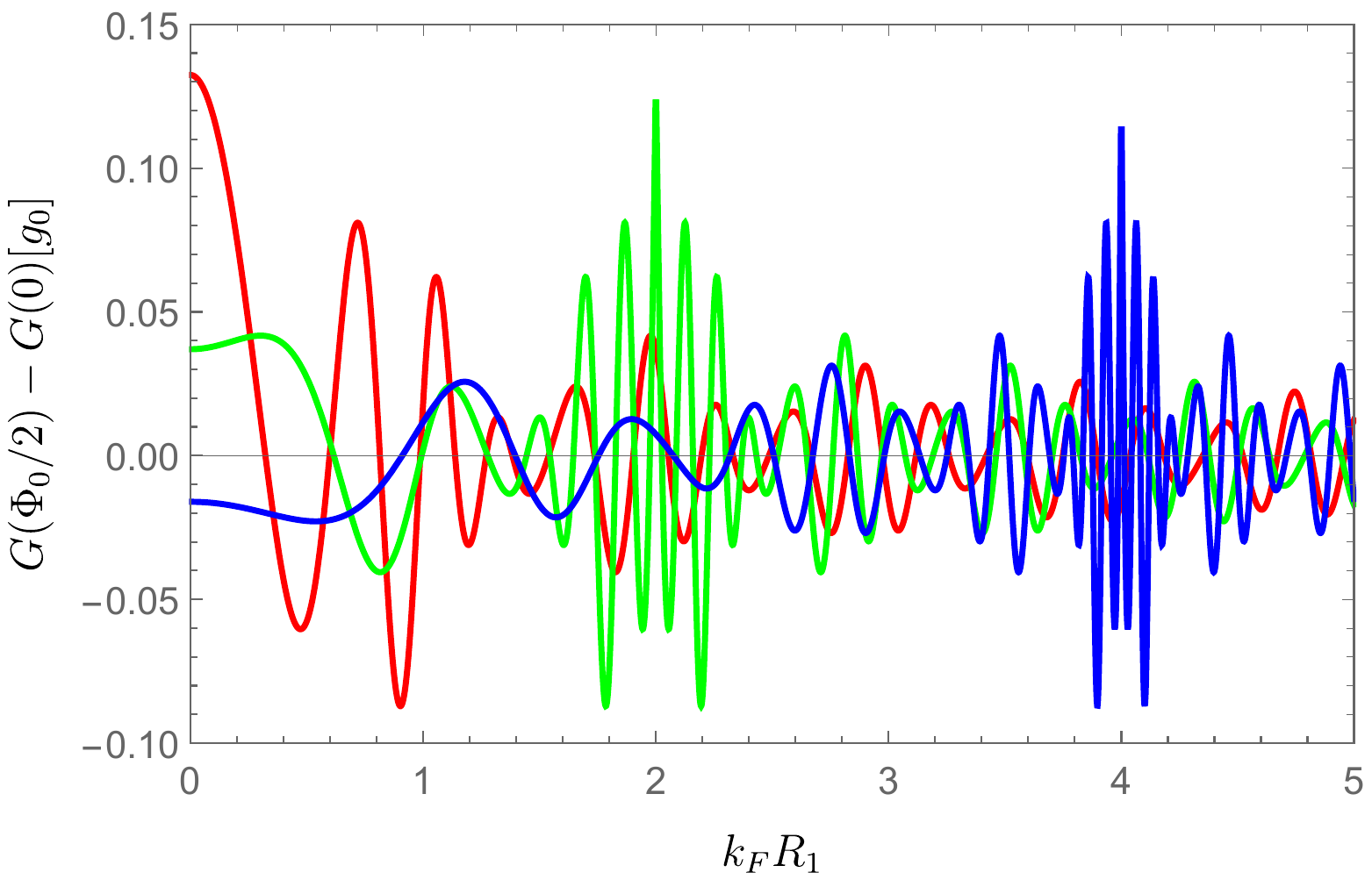}
	\includegraphics[width=0.32\linewidth]{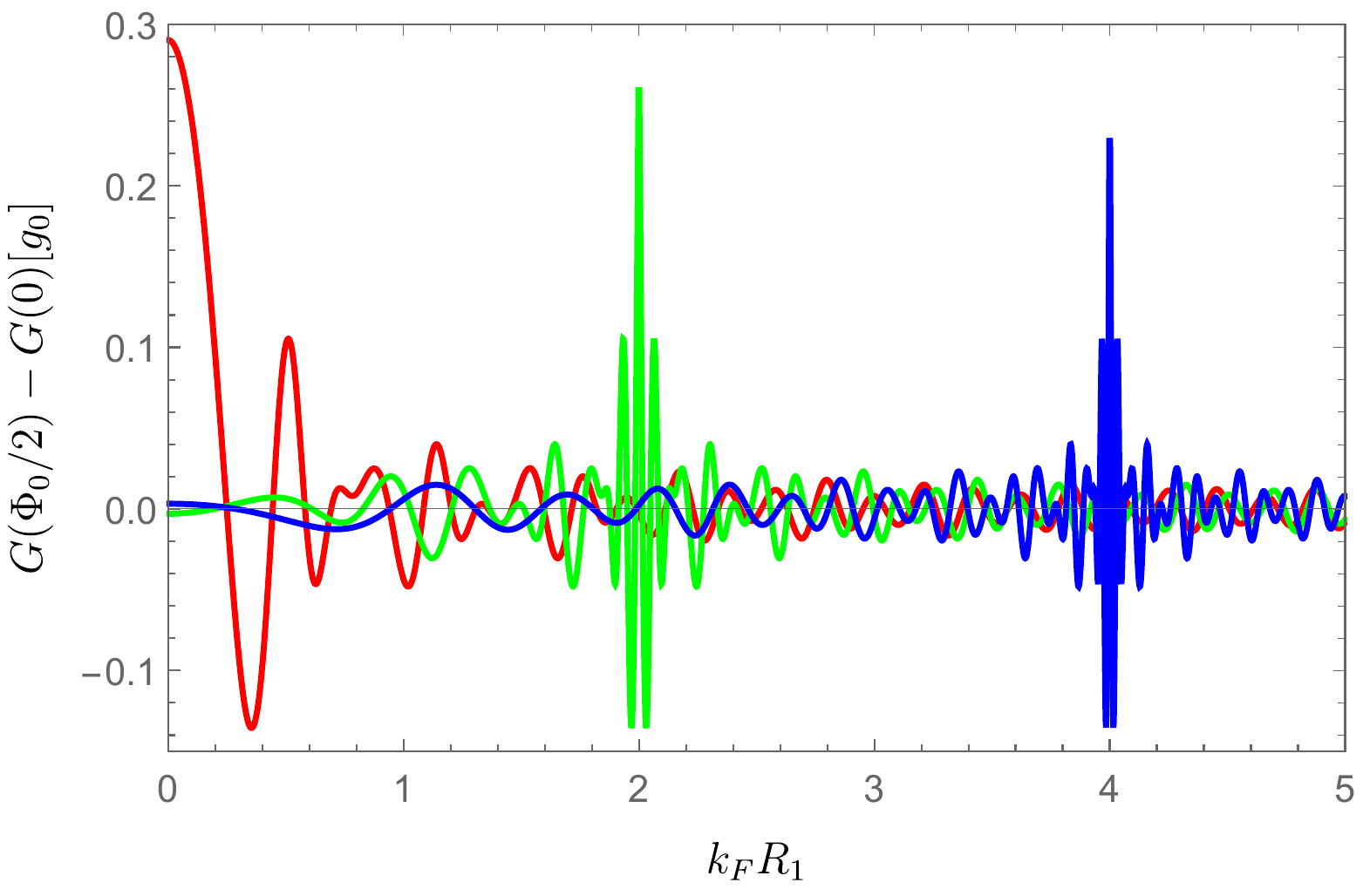}
	\caption{\sf (color online) The magnitude of the conductance oscillations $ \Delta G = G(\Phi_0/2) - G(0)) $
		displayed as a function of the doping $k_F R_1$ for  different values of $ R_1 \delta = 0 $ (red line), $ 2 $ (green line), $ 4 $ (blue line) with  the radii ratio  $R_2/R_1$: 2.5 (left panel), $ 5 $ (middle panel), $ 7.5 $ (right panel).}
	\label{fig 8}
\end{figure}
We consider now  the magnitude of the conductance oscillations $\Delta G$ as being  the difference between $G(\Phi_0/2)$ and $G(0)$ 
\begin{equation}\label{eq25}
\Delta G= G(\Phi_0/2)-G(0)
\end{equation}
which is presented 
as a function of the doping $ k_F R_1 $ in Fig. \ref{fig 8}. In tunnel mode and for different doping values (close to the neutral point and even with high doping), the magnitude of the conductance oscillations (\ref{eq25}) takes relatively large values ($ \Delta G>0.1 g_0 $) for moderate radii ratio $ R_2/R_1\geq5 $. This difference is valid in the case where $ k_F R_1 \approx\nu$. 
In the presence of energy gap, we observe an increase in $ \Delta G $ with zero doping $ k_F R_1 = 0 $ for small radii ratio and it disappears with its increase. In the middle panel where $ R_2 / R_1 = 5 $ and in comparison with $ \Delta G(R_1 \delta = 0) $ (red line)  \cite{Rycerz2020} we observe the appearance of a resonance peak corresponding to the values chosen for of the energy gap. The frequency of these resonance peaks increases when the energy gap increases or when we increase the radii ratio (see the right panel $ R_2 / R_1 = 7.5 $). It is also important to notice that the distance between two successive nodes in a series of discrete doping values for which $ \Delta G=0 $ decreases as one approaches the peak, i.e. the sign alternation of $ \Delta G$ becomes very fast in the vicinity of these peaks. To give an illustration, we consider for example $ R_2/R_1=5 $ and then the first five nodes of $ \Delta G=0 $ for $ k_F R_1=2 $ (green line) correspond to the following values
\begin{equation}\label{eq26}
(k_F)_{\Delta G=0}=0.615,\quad0.994,\quad1.277,\quad1.445,\quad1.544
\end{equation}
and for $ k_F R_1 = 4 $ (blue line) we have 
\begin{equation}\label{eq27}
	(k_F)_{\Delta G=0}=0.909,\quad1.394,\quad1.752,\quad2.072,\quad2.318
\end{equation}

\begin{figure}[H]\centering
	\includegraphics[width=0.24\linewidth]{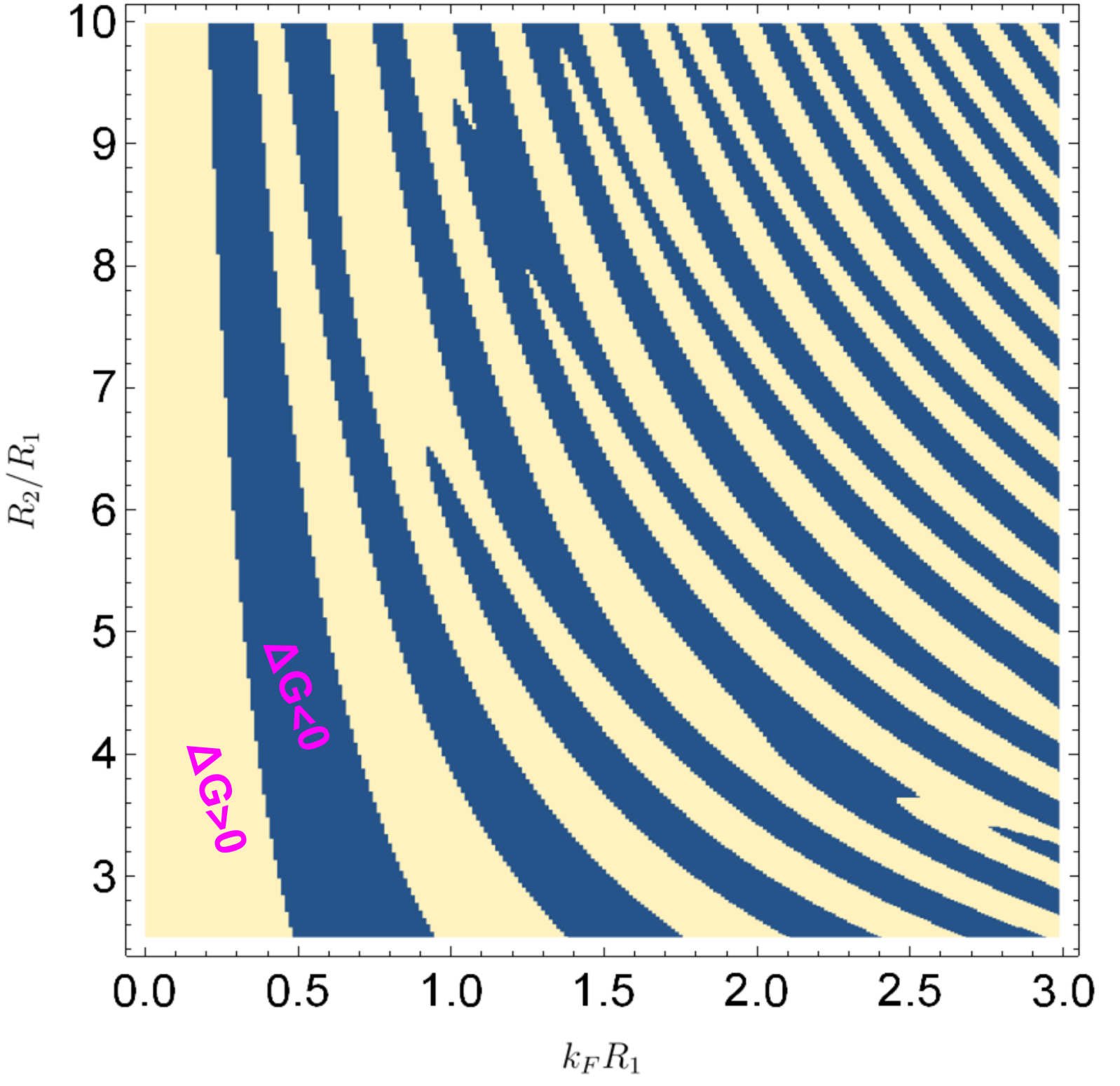}
	\includegraphics[width=0.24\linewidth]{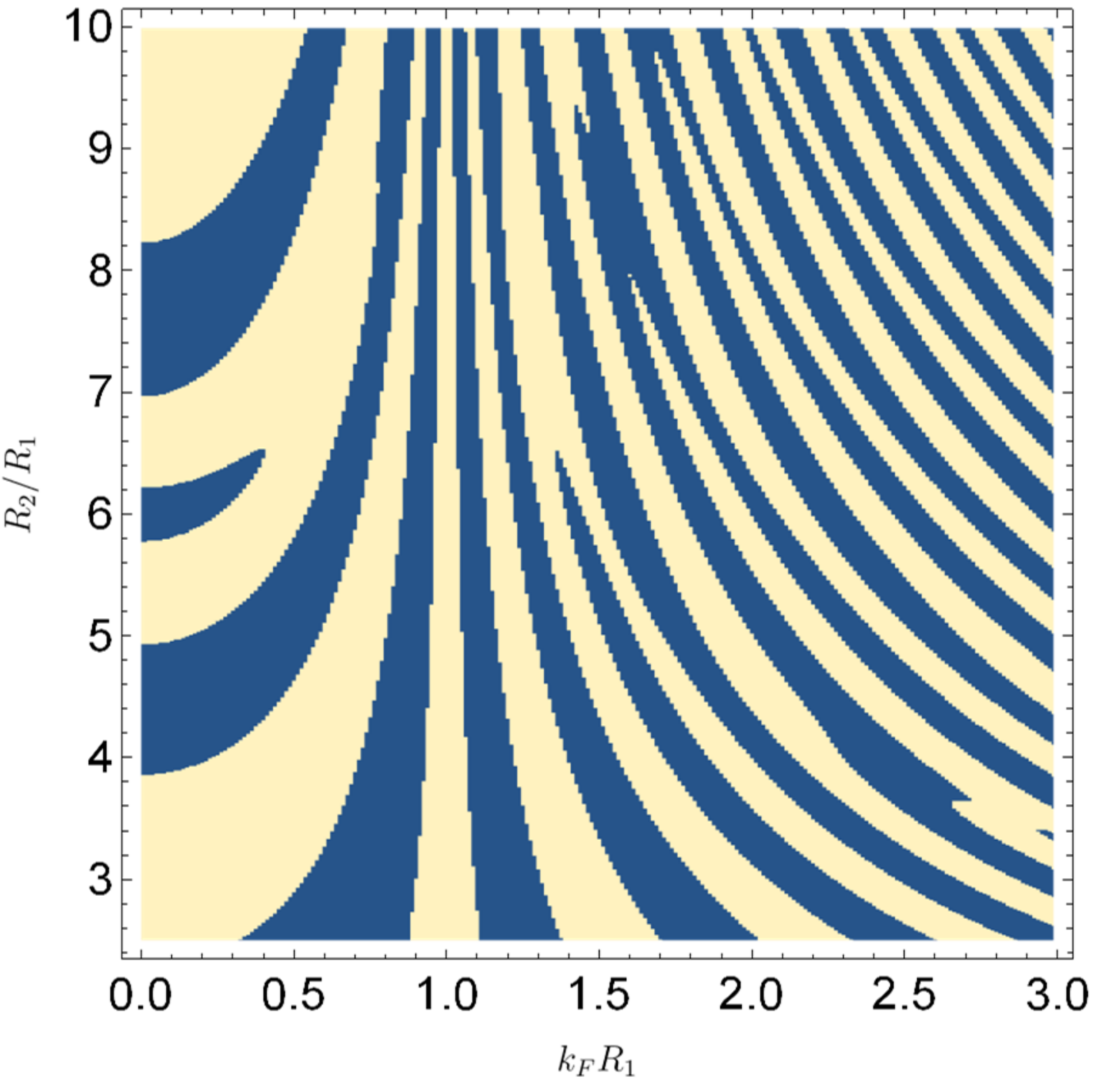}
	\includegraphics[width=0.24\linewidth]{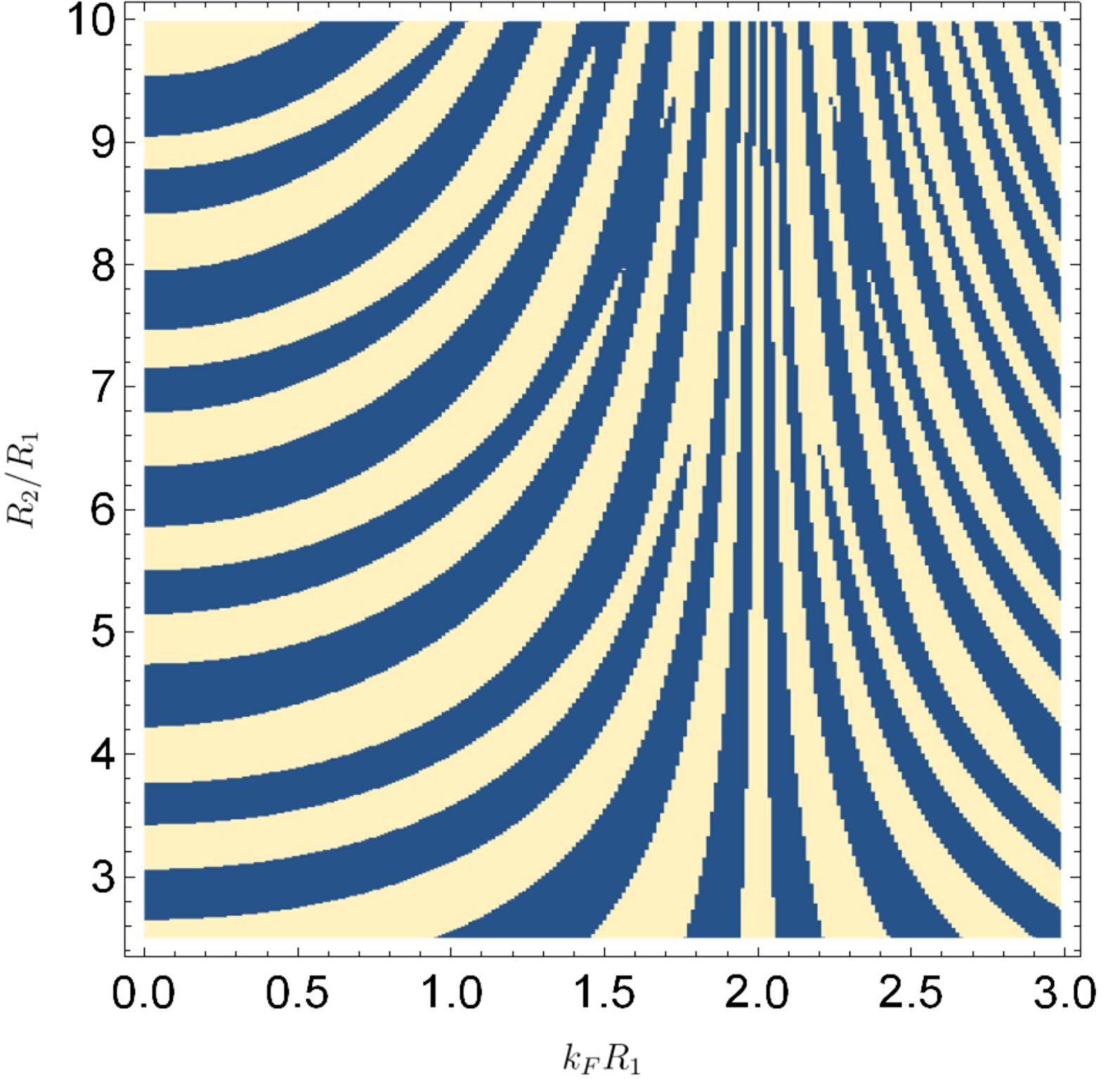}
	\includegraphics[width=0.24\linewidth]{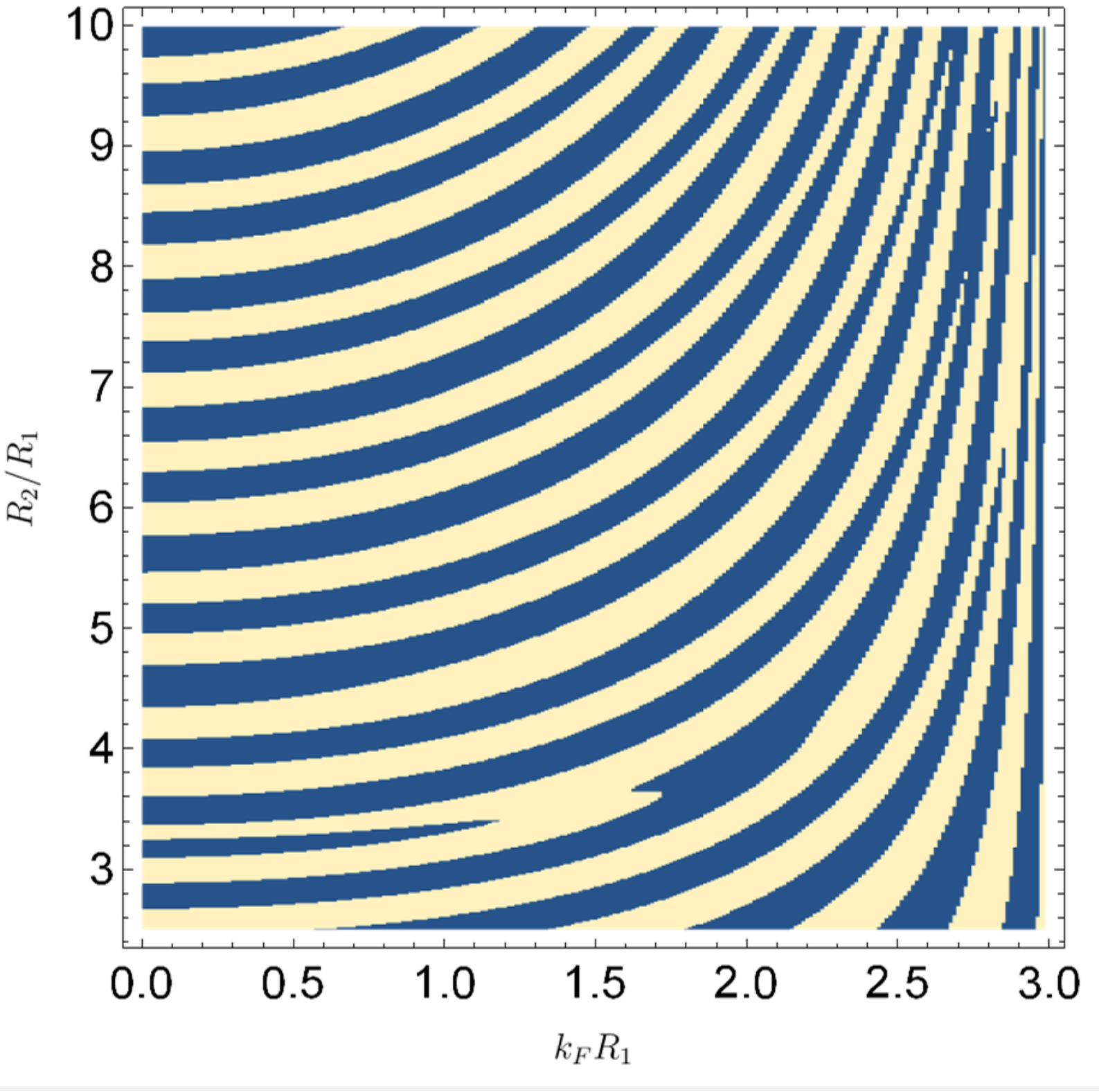}
	\caption{\sf (color online) Nodal lines of $ \Delta G $ separated by areas with $ \Delta G > 0 $ (beige) and $ \Delta G < 0 $ (blue) as a function of the doping $ k_F R_1 $, radii ratio $ R_2/R_1 $ and different values of energy gap $ R_1 \delta = 0,  1,  2,  3 $ respectively.}
	\label{fig 9}
\end{figure} 
Fig. \ref{fig 9} shows the nodal lines of $ \Delta G $ as a function of doping and radius ratio separated by areas with $ \Delta G>0 $ (beige) and $ \Delta G<0 $ (blue) and under the effect of four values of energy gap  $ R_1 \delta = 0,  1,  2, 3 $. We observe a reduction of the patterns and a tilting of the nodal lines in the vicinity of the energy gap values indicating the presence of the resonances observed in Fig. $\ref{fig 8}$.

\begin{figure}[H]\centering
	\includegraphics[width=0.33\linewidth]{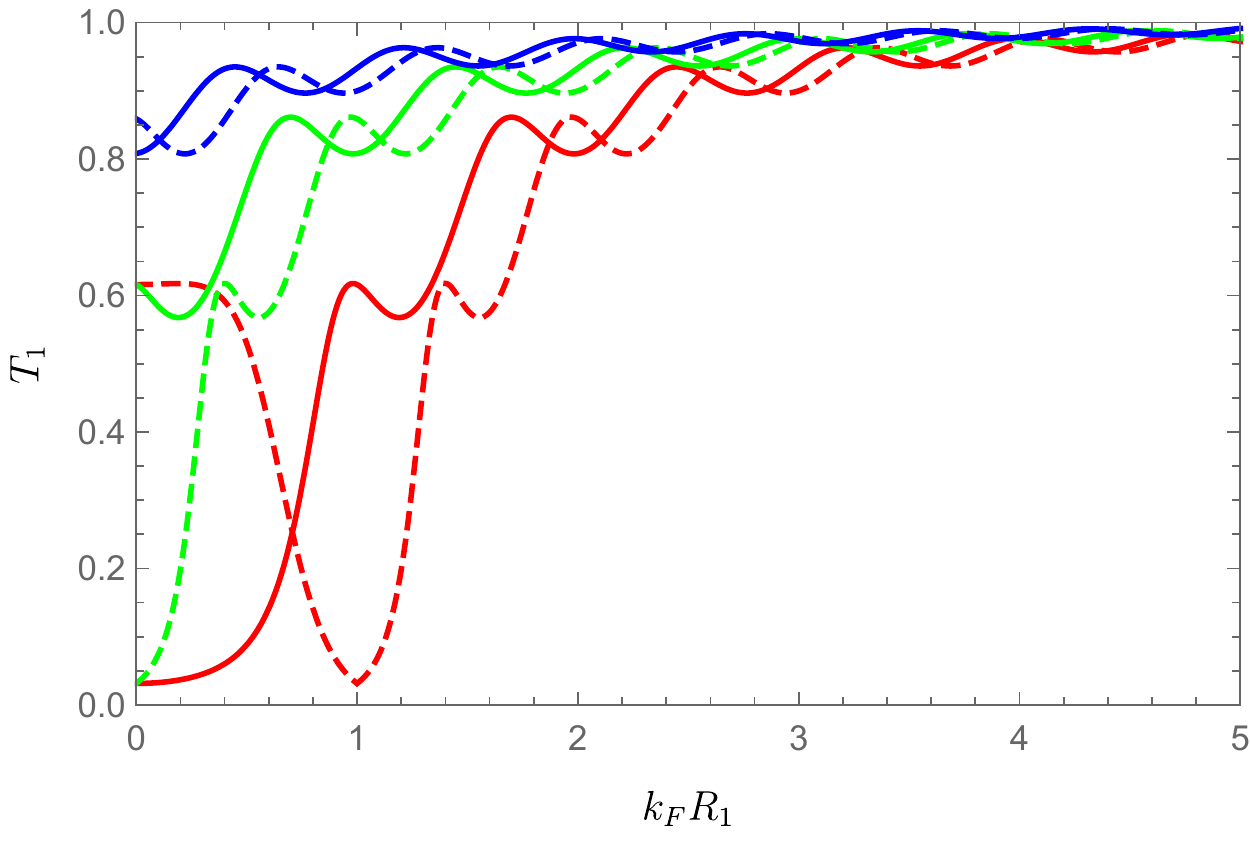}\ \ \ \ \ \ \
	\includegraphics[width=0.33\linewidth]{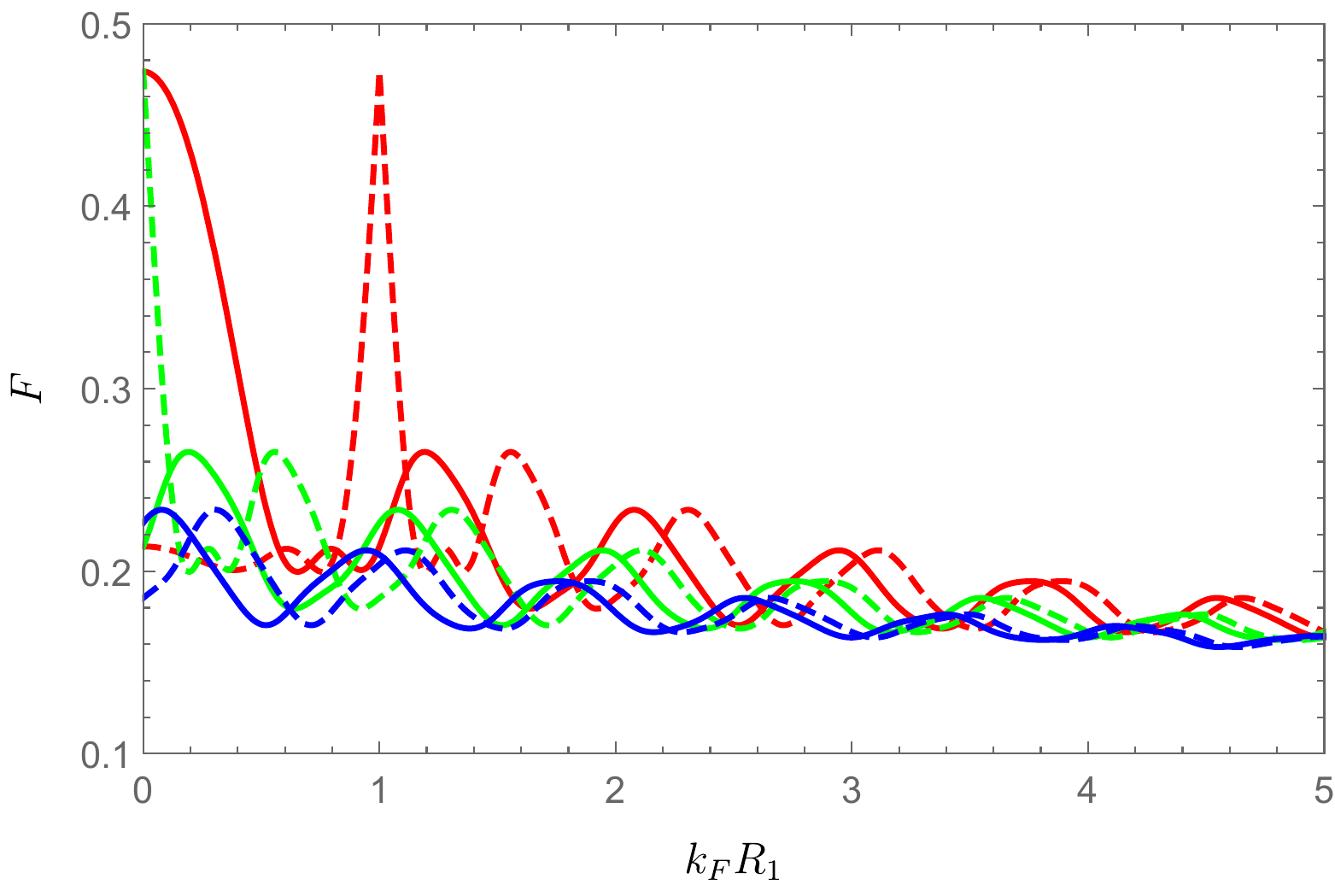}\\
	\includegraphics[width=0.33\linewidth]{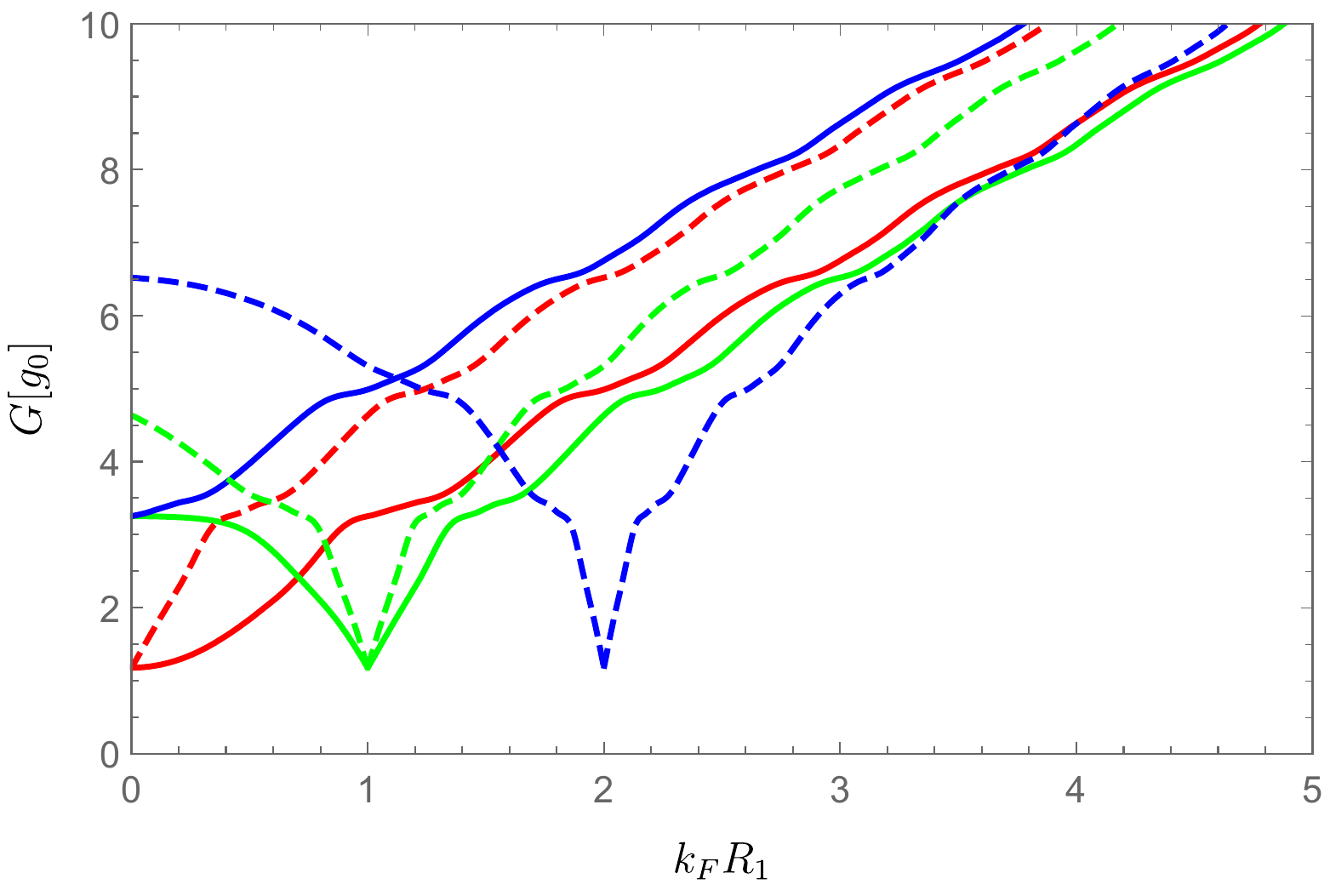}\ \ \ \ \ \ \
	\includegraphics[width=0.33\linewidth]{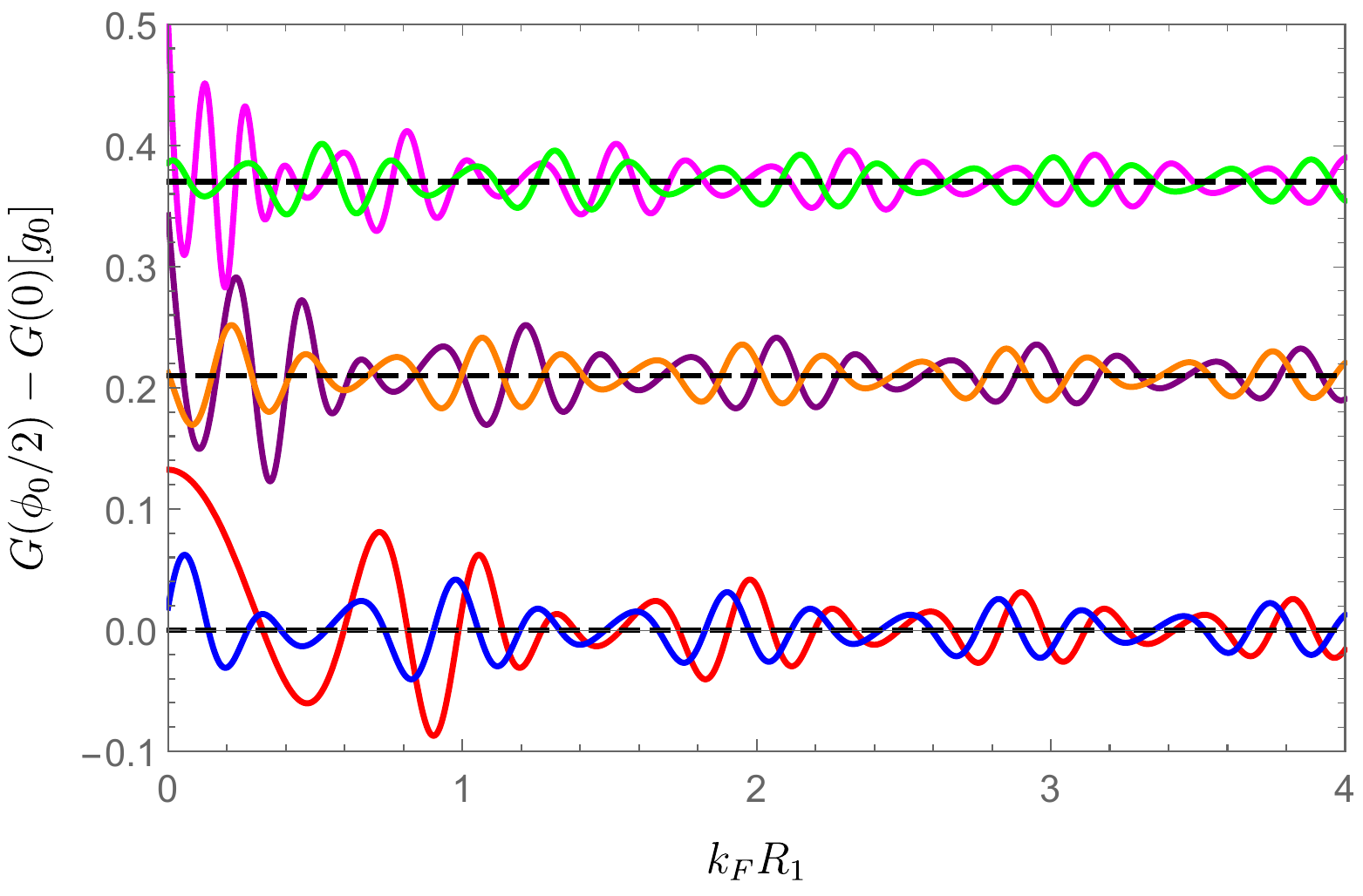}\\
	\caption{\sf (color online) The transmission $T_1$,  Fano factor $F$,  conductance $G$ and  magnitude of the conductance oscillations $\Delta G$ as function of the doping $k_F R_1$ for the radii ratio  $ R_2/R_1=5 $ and different values of  energy gap and electrostatic potential:
		$(R_1\delta=0$, $ R_1 u_0=0$) (red line), $(1, 0)$ (green line), $ (0,1) $ (blue line), $ (1,1) $ (red dashed line), $ (2,1) $ (green dashed line), $ (3,1) $ (blue dashed line). 
		For $\Delta G$ we choose
		($R_1 \delta =0$, $R_1 u_0=0$) (red line), ($0,1$) (blue line), ($1,1$) (purple line), ($1,2$) (orange line), ($2,2$) (magenta line), ($2,3$) (green line). }
	\label{fig 10}
\end{figure} 

We now add an electrostatic potential term $ U $ and investigate its effect and  that of the energy gap in Fig. \ref{fig 10} showing the transmission $ T_1 $,  Fano factor $ F $,  conductance $ G $ and  magnitude of the conductance oscillations $ G $ as a function of the doping $ k_F R_1 $. The effect of the electrostatic potential at zero gap is marked by an increase in transmission and conductance. The presence of the energy gap reduces the effect of the potential if $ R_1u_0 \geqslant R_1\delta $ (first and second panel on the left). The Fano factor becomes minimal and the potential eliminates the peaks created by the gap (first panel on the right). In the last panel, the comparison at zero energy gap of a potential $ R_1 u_0 = 0 $ (red line) and $ R_1 u_0 = 1 $ (blue line) shows the increase of the sign $ \Delta G $ alternation rate. For the case $ R_1 \delta = 1 $ and the potential varies from $ 1 $ to $ 2 $, we observe a coincidence of periods followed by a decrease of amplitudes for $ k_F \leqslant 0.52 $.

\section{Conclusion}

We have studied the effect of an energy gap created in the area bounded by the inner and outer radii of a Corbino disk in single-layer graphene  pierced by a long solenoid creating a current $ I_s $ generating on its part a magnetic flux $\Phi_i$, in the presence and absence of an electrostatic potential $ U $. Taking advantage of the geometry of the Corbino disk, we have performed theoretical studies using mode matching based on the effective Dirac equation. Thus we determined the transmission probability of an electron of given angular momentum crossing the Corbino disk in graphene and subsequently the associated conductance as well as Fano factor.

The effect of the energy gap on the parameters of our system is illustrated as follows. An increase of the transmission at zero doping, suppression of the tunneling effect at the points $ k_F R_1=R_1 \delta $ and an oscillatory aspect of the transmission  as a function of the radii ratio $ R_2/R_1 $. A coincidence of the periods followed by a decrease of the amplitudes of the 
conductance, then a displacement around the value $ G(R_1\delta=0) $ following the sign of the difference $ k_F R_1-R_1\delta $. An appearance of resonance peaks of magnitude of the conductance oscillations $ \Delta G $, followed by an increase in the alternation speed of its sign in the vicinity of the points $ k_F R_1=R_1 \delta $. Finally the electrostatic potential breaks the symmetry and allows to control the effect of the energy gap for the cases where $ R_1 u_0\geqslant R_1\delta $.   

\end{document}